\documentclass[mcmahonlabnew, oneside, onecolumn, Numbered]{jnl}

\usepackage{graphicx}
\usepackage{multirow}
\usepackage{amsmath,amssymb,amsfonts}
\usepackage{amsthm}
\usepackage{mathrsfs}
\usepackage[title]{appendix}
\usepackage{xcolor}
\usepackage{textcomp}
\usepackage{manyfoot}
\usepackage{booktabs}
\usepackage{anyfontsize}
\usepackage{braket}
\usepackage{siunitx}
\usepackage[ruled,vlined,linesnumbered]{algorithm2e}
\SetKwInput{KwInput}{Input}
\SetKwInput{KwOutput}{Output}

\usepackage{accents}
\usepackage{bibunits}
\usepackage{titletoc}

\defaultbibliographystyle{mcmahonlabnew}
\hypersetup{urlcolor=black}

\def\edit#1{{ {#1}}}

\title{
Microwave signal processing using an analog quantum reservoir computer
}
\author[1,2]{\fnm{Alen} \sur{Senanian}}
\author[1,2]{\fnm{Sridhar} \sur{Prabhu}}
\author[2]{\fnm{Vladimir}\sur{Kremenetski}}
\author[1,2]{\fnm{Saswata} \sur{Roy}}
\author[3,4]{\fnm{Yingkang} \sur{Cao}}
\author[2]{\fnm{Jeremy} \sur{Kline}}\presentaddress{%
 \footnotesize \orgdiv{Department of Electrical Engineering and Computer Science},
 \orgname{Massachusetts Institute of Technology},
 \orgaddress{%
   \state{MA}%
   , \country{USA}}%
   \\  $^{\dagger}$\fontsize{8}{8}\selectfont Present address: %
 \footnotesize \orgdiv{Department of Applied Physics},
  \orgname{Yale University},%
 \orgaddress{
 \state{ CT}%
  , \country{USA}}%
   }
\author[2,5]{\fnm{Tatsuhiro} \sur{Onodera}}
\author[2,5]{\fnm{Logan~G.} \sur{Wright}}
\presentaddressy{%
 \footnotesize \orgdiv{Department of Electrical Engineering and Computer Science},
 \orgname{Massachusetts Institute of Technology},
 \orgaddress{%
   \state{MA}%
   , \country{USA}}%
   \\  $^{\dagger}$\fontsize{24}{24}Present address: %
 \footnotesize \orgdiv{Department of Applied Physics},%
 \orgname{Yale University},%
 \orgaddress{%
   \state{ CT}%
   , \country{USA}}}
\author[3,4]{\fnm{Xiaodi} \sur{Wu}}
\author[2]{\fnm{Valla} \sur{Fatemi}}
\author[2,6]{\fnm{Peter~L.}~\sur{McMahon}}

\affil[1]{\orgdiv{Department of Physics}, \orgname{Cornell University}, \state{NY}, \country{USA}}
\affil[2]{\orgdiv{School of Applied and Engineering Physics}, \orgname{Cornell University},   \state{NY}, \country{USA}}
\affil[3]{%
  \orgdiv{Department of Computer Science}, \orgname{University of Maryland},  \state{MD}, \country{USA}}
\affil[4]{%
  \orgdiv{Joint Center for Quantum Information and Computer Science}, \orgname{University of Maryland},  \state{MD}, \country{USA}}
\affil[5]{%
 \orgdiv{NTT Physics and Informatics Laboratories}, \orgname{NTT Research, Inc.}, \state{CA}, \country{USA}}
\affil[6]{%
  \orgdiv{Kavli Institute at Cornell for Nanoscale Science}, \orgname{Cornell University},  \state{NY}, \country{USA}}

\begin{document}
\begin{bibunit}

\abstract{
Quantum reservoir computing (QRC) has been proposed as a paradigm for performing machine learning with quantum processors where the training is efficient in the number of required runs of the quantum processor and takes place in the classical domain, avoiding the issue of barren plateaus in parameterized-circuit quantum neural networks. It is natural to consider using a quantum processor based on superconducting circuits to classify microwave signals that are \textit{analog}---continuous in time. However, while theoretical proposals of analog QRC exist, to date QRC has been implemented using circuit-model quantum systems---imposing a discretization of the incoming signal in time, with each time point input by executing a gate operation. In this paper we show how a quantum superconducting circuit comprising an oscillator coupled to a qubit can be used as an analog quantum reservoir for a variety of classification tasks, achieving high accuracy on all of them. Our quantum system was operated without artificially discretizing the input data, directly taking in microwave signals. Our work does not attempt to address the question of whether QRCs could provide a quantum computational advantage in classifying pre-recorded classical signals. However, beyond illustrating that sophisticated tasks can be performed with a modest-size quantum system and inexpensive training, our work opens up the possibility of achieving a different kind of advantage than a purely computational advantage: superconducting circuits can act as extremely sensitive detectors of microwave photons; our work demonstrates processing of ultra-low-power microwave signals in our superconducting circuit, and by combining sensitive detection with QRC processing within the same system, one could achieve a quantum sensing-computational advantage, i.e., an advantage in the overall analysis of microwave signals comprising just a few photons. 
}

\phantomsection
\addcontentsline{toc}{section}{Abstract}

\maketitle

\phantomsection
\addcontentsline{toc}{section}{Introduction}

\section*{Introduction}\label{sec:Introduction}

Over the last decade, researchers in quantum information processing have broadly divided their efforts into two distinct but complementary directions. In one, the focus has been on realizing the building blocks for large-scale, fault-tolerant quantum processors~\cite{ladd2010quantum,jones2012layered,campbell2017roads}, which would enable running algorithms such as Shor's or Grover's at meaningful scale. In the other, there has been a push to realize quantum systems comprising tens to hundreds of qubits or qumodes, but without error correction, and to explore what can be done with such noisy, pre-fault-tolerance systems---often denoted as noisy, intermediate-scale, quantum (NISQ) devices~\cite{Preskill_2018}. Quantum computational supremacy with such NISQ devices has been demonstrated~\cite{arute2019quantum,Zhong_2020}, but there has been much less progress on achieving quantum advantage in practically relevant applications~\cite{mohseni2017commercialize}. There have been many NISQ studies on quantum machine learning~\cite{biamonte2017quantum}, and in this area too, quantum advantage for problems of broad practical interest has remained elusive~\cite{schuld2022quantum,cerezo2022challenges}. A key challenge in quantum neural networks realized with parameterized quantum circuits has been training the parameters when the optimization landscape suffers from barren plateaus~\cite{mcclean2018barren, wang2021noise, marrero2021entanglement, arrasmith2022equivalence}. A major open question is whether one can achieve any practically relevant advantage for machine learning with NISQ systems.

\begin{figure*}[h]
    \centering
    \includegraphics[width=0.8\textwidth]{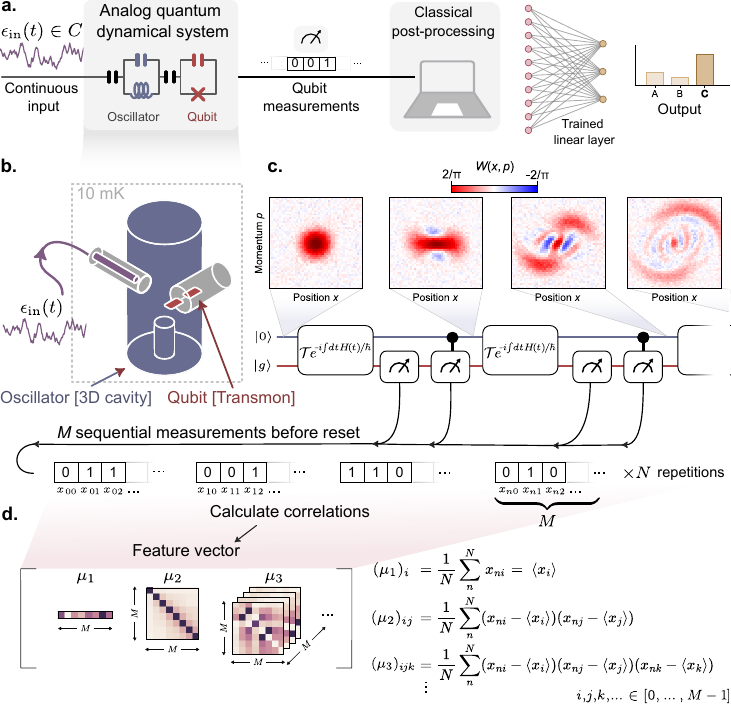}
    \caption{
    \textbf{Analog signal classification with a continuous-variable quantum reservoir computer (QRC) using measurement trajectories.}
    \textbf{(a.)} 
        We perform machine learning using a quantum system consisting of an oscillator coupled to a qubit. Analog signals are fed into our analog quantum dynamical system, continually displacing the oscillator mode while the qubit is projectively measured. The measurement trajectories provide complex features a digital linear layer can use to perform classification on a variety of tasks. 
    \textbf{(b.)}
        The signals interface directly with the qubit-oscillator system, composed of a 3D aluminum cavity (blue) hosting a transmon qubit (red) and a readout resonator (omitted for clarity; see Appendix~\ref{sec:si:experimental_setup}).
    \textbf{(c.)}
        Wigner tomography performed on the oscillator state through various stages of the reservoir dynamics. The dynamics include entanglement-generating unitary evolution, and projective measurements of both the qubit and oscillator. The back-action produced by the measurements add to the non-classical features generated by the entangling unitaries. The balance of measurements and unitaries, which do not commute with each other in our implementation, lead to complex correlations in the measurement trajectories. 
    \textbf{(d.)} 
        The digital linear layer performs classification based on a feature vector, which we construct using the expectation values of the central moments $\mu_p$ ($p = 1,2,3,\ldots$), which capture the essential correlations in the reservoir dynamics. 
    }
    \label{Fig:Main}
\end{figure*}

Inspired by the framework of reservoir computing \cite{lukovsevivcius2009reservoir, schrauwen2007overview, khan2021physical, Gauthier_2021} in classical machine learning, quantum reservoir computing (QRC) \cite{Fujii2021, fujii2017harnessing, Ghosh_2019,PhysRevResearch.3.013077, PhysRevResearch.4.033007,Spagnolo2022} has emerged as an approach to quantum machine learning that entirely avoids barren plateaus by performing all learning in a final, linear layer. They key idea of a QRC is that a quantum system (called a \emph{quantum reservoir}) can generate nonlinear, high-dimensional features of inputs to it, and that these features can be used to perform machine-learning tasks purely by training a classical linear transformation. However, experimental demonstrations to date have been performed with digital quantum circuits \cite{Pfeffer_2022, PhysRevApplied.14.024065, kubota2022quantum, mlika2023user, yasuda2023quantum, Suzuki_2022, PhysRevX.13.041020} that have limited the complexity of tasks that can be performed, in part due to an input bottleneck imposed by the use of discrete gates to input temporal data using a series of separate, imperfect gates.

The aim of our work is to demonstrate a proof-of-principle for a new application of and approach to quantum machine learning with NISQ devices that overcomes or sidesteps the challenges in training and inputs noted above. We use the driven, continuous-time analog quantum nonlinear dynamics of a superconducting microwave circuit as a quantum reservoir to generate features for classifying weak, analog microwave signals (Fig.~\ref{Fig:Main}a). We use repeated measurements of the reservoir both to extract features that contain information about temporal correlations in the input data, as well as to induce non-unitary dynamics. Our use of a continuous-variable system in our quantum reservoir grants us access to a substantially larger Hilbert space than would be the case with a qubit-only system with equally many hardware components. Our approach is similar to proposals for analog NISQ processors and simulators \cite{ParraRodriguez2020, garcamolina2021noise, daley2022practical}, which aim to avoid the overhead caused by imposing a discrete-time abstraction. Analog operation grants us an even more important ability however, which fundamentally distinguishes our work from prior experimental demonstrations of quantum machine learning on circuit-model quantum processors: it allows our device to directly, natively receive weak analog microwave signals, and to immediately leverage analog quantum information processing to extract relevant features of the signals for classification.

Rather than focusing on using NISQ hardware to perform computation on pre-recorded, digital data, we instead use quantum hardware to perform computation on real-time analog signals that interface directly with our microwave superconducting device. Our experiments do not address the question of whether a QRC can achieve a quantum computational advantage, since our experimental device is small enough to be easily classically simulable. However, our demonstrations suggest a route to achieving a quantum advantage of a different kind: an advantage in the quantum detection and processing of weak microwave signals, allowing quantum hardware to extract complex information of interest from dim, analog signals in ways that would be noisier with a conventional classical approach. This type of quantum advantage, arising from a combination of quantum sensing with extraction of complex features about the sensed signal, is discussed in general terms as a route to quantum advantage with quantum machine learning in Ref.~\cite{cerezo2022challenges}. Our work shows that when classical signals comprising just a few photons have entered an analog quantum reservoir, they can be classified using our QRC approach. The signals we classify are synthesized at room temperature and pass through 60 dB of attenuation before reaching our device. However, if instead one combines this analog quantum processing with a sensitive quantum detector of microwave radiation, as has already been previously demonstrated using superconducting circuits \cite{wang2021quantum, wang2022quantum, backes2021quantum, assouly2023quantum, Dixit2021}, then one can construct a system that achieves a quantum advantage in the task of combined sensing and signal processing of high temperature signals.

\subsection*{Experimental setup and protocols}\label{sec:experiment}
\addcontentsline{toc}{subsection}{\nameref{sec:experiment}}

Our quantum reservoir, composed of a cavity resonator coupled to a transmon (Fig.~\ref{Fig:Main}b), can be modeled with the following qubit-oscillator Hamiltonian in the rotating-frame,
\begin{equation}
    H/\hbar = -\chi \vert e \rangle \langle e \vert a^{\dagger}a  + \epsilon(t)a^{\dagger} +  \Omega(t) \vert e \rangle \langle g \vert + \mathrm{H.c.},
    \label{eq:Hamiltonian}
\end{equation}
where $\vert g \rangle$ and $\vert e \rangle$ define the qubit subspace of the transmon, $a$ is the photon annihilation operator of the oscillator mode, and $\chi$ is the nonlinear interaction strength (see Appendix~\ref{sec:si:reservoir:hamiltonian} for details). The third term of Eq.~\ref{eq:Hamiltonian} describes the unitary control of the qubit with a time-dependent drive $\Omega(t)$, and the second term describes both the encoding of the input data $\epsilon_{\mathrm{in}}(t)$, and unitary control of the oscillator mode, i.e., $\epsilon(t) = \epsilon_{\mathrm{in}}(t) + \epsilon_{\mathrm{control}}(t)$. Equation~\ref{eq:Hamiltonian} describes the unitary dynamics, which is complemented by non-unitary dynamics generated by the back-action from qubit measurements interspersed throughout the evolution.

The oscillator and qubit control drives used in this paper realize a reservoir that consists of a series of entangling unitaries interleaved with qubit and oscillator measurements (Fig.~\ref{Fig:Main}c). The analog input is sent resonantly to the cavity and results in a time varying conditional displacement of the oscillator, which streams in concurrently with control drives. The cavity resonator hosting the oscillator mode has a resonance frequency of \SI{6}{\giga\hertz} and a 2-\SI{}{\kilo \hertz} linewidth. The combination of the input and control drives implement a unitary that encodes the input into the state of the oscillator and generates entanglement between the qubit and the oscillator. Following the unitary evolution, we perform a qubit measurement, and then a parity measurement of the oscillator state~\cite{PhysRevX.10.021060, Heeres_2015} (see Appendix~\ref{sec:si:reservoir_characterization:measurements}). The parity measurement projects the oscillator state into super-positions of either even or odd Fock states, a highly non-Gaussian measurement allowing one to sense changes in the photon-number distribution. Additionally, the entangling dynamics between the measurements effectively implement a sequence of non-commuting measurements (see Appendix~\ref{sec:si:reservoir:time_independent}), generating correlated measurement distributions that can then be used as complex output features. Finally, after four rounds of applying the unitary and the qubit-oscillator measurements, we reset the system before repeating the scheme so that we may collect many samples of the measurement trajectory. The reset, which occurs at a rate much faster than the decoherence rate of the oscillator, additionally ensures that our system remains coherent.

\begin{figure*}[h]
\centering
    \includegraphics[width=0.68\textwidth]{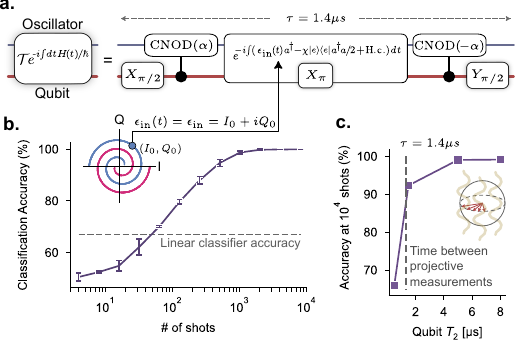}
    \caption{
    \textbf{Reservoir protocol overview with an example time-independent classification task}
    \textbf{(a)}
        The unitary dynamics in our reservoir are generated by control pulses that serve to entangle the qubit with the oscillator before the analog input is received by the oscillator. For tasks where the analog data is time-independent, the dynamics are fully gate-based, and the oscillator is dis-entangled with the qubit before the qubit and oscillator measurements. For details of the motivation behind the particular unitaries implemented for our reservoir, see Appendix~\ref{sec:si:reservoir}.
    \textbf{(b)}
        (Inset) An illustrative machine learning example is the classification of time-independent signals from two arms of a Spiral distribution defined in the signal $I-Q$ plane.  For quantum machine learning, unlike classical, the performance is unavoidably impacted by sampling noise. Here, we plot the classification accuracy of the spiral task against an increasing number of shots. Also plotted is the performance of a linear layer acting directly on the two-dimensional $I,Q$ data, indicating that non-linearity is required to perform this task with sufficient accuracy. 
    \textbf{(c)}
        Classification accuracy at $10^4$ shots as a function of qubit coherence time that we tune via resonator-induced dephasing during the classification (see Appendix~\ref{sec:si:reservoir_characterization:t2}). While we see a large drop in classification performance when the qubit coherence time is heavily suppressed and the system is completely disentangled, the performance only begins to suffer once the qubit $T_2$ approaches the duration between measurements. 
    }
    \label{Fig:Time-Independent}
\end{figure*}

The measurement outcomes are used to construct output feature vectors to be fed into the linear layer, but this can be done in a few different ways. When performing repeated measurements on our system, we generate a sample bitstring of length $M$ describing the quantum trajectory over $M$ measurements. After $M$ measurements are performed, we reset the system and repeat the procedure, each time generating a bitstring $\vec{x}_n = [x_{n0}, x_{n1}, \ldots, x_{nM-1}]$, where $n$ refers to the $n$th sample (Fig.~\ref{Fig:Main}c). The outcomes can be counted to directly form a sample probability distribution $p(\vec{x}\vert \epsilon_{\mathrm{in}}(t))$ over measurement trajectories, which can then be used as a high-dimensional output feature vector after obtaining a sufficient number of samples $N$. While this approach has the benefit of capturing all information in the measurement distribution~\cite{PhysRevX.13.041020}, it can generally suffer from poor scaling in sampling noise, requiring $N \sim 2^{M}$ shots in the worst case \cite{wright2019capacity}. 

Here, we construct an output feature vector from estimates of successive central moments $\mu_1, \mu_2, \mu_3, \ldots$ of the underlying distribution $p(\vec{x}\vert \epsilon_{\mathrm{in}}(t))$ (Fig.~\ref{Fig:Main}d). For example, the first-order central moment $\mu_1$ is a $M$-dimensional vector representing the average over all measured bitstrings, i.e. $\mu_1 = [\langle x_{n0} \rangle , \langle x_{n1} \rangle ,\ldots ]$, the second-order central moment $\mu_2$ is the covariance matrix with elements $(\mu_2)_{ij} = \langle x_{ni}x_{nj} \rangle - \langle x_{ni}\rangle \langle x_{nj} \rangle$, and so on. Here, the expectation value is taken over the sample index $n$. This approach, inspired by Ref.~\cite{khan2021physical}, has the benefit of leveraging the hierarchy of noise in the central moments, while capturing the essential correlations in the dynamics to achieve high accuracy even in the few-sample regime. Furthermore, the output feature vector dimension only scales polynomially with the number of measurements, where the highest polynomial power is given by the order of the highest central moment, which we restrict to 3 for all tasks in this work. Finally, given finite memory in our reservoir, we further restrict the output vector by choosing to only calculate correlations between measurements at most 3 measurements apart. These truncated moments are then flattened and concatenated to construct our output feature vectors. In all, for the $M$ = 8 measurements we use in this work, the resultant output feature vector size with this prescription is 94. For a detailed discussion of the construction of our reservoir output features with comparisons of different encodings, see Appendix~\ref{sec:si:ML:output_feature_encoding}.

\section*{Results}\label{sec:results}
\addcontentsline{toc}{section}{\nameref{sec:results}}

\subsection*{Classification of time-independent signals}\label{sec:results:spiral}
\addcontentsline{toc}{subsection}{\nameref{sec:results:spiral}}

To illustrate the scheme proposed in this work, we begin with an example classification using our quantum reservoir by performing binary classification task of time-independent signals. Figure~\ref{Fig:Time-Independent}a describes the control drives in more detail. For time-independent input data, the two-dimensional input data is encoded as the $I$ and $Q$ quadratures of an analog signal resonant with the cavity resonance frequency. In the rotating frame of the system (Eq.~\ref{eq:Hamiltonian}), this is effectively a time-independent signal, i.e. $\epsilon_{\mathrm{in}}(t) = \epsilon_{\mathrm{in}} = I+Q$, which results in a displacement of the oscillator state. For such time-independent tasks, the signal bandwidth is set by its duration which, in general, can make the resultant displacement conditioned on the qubit state due to the cross-Kerr interaction (see first term of Eq.~\ref{eq:Hamiltonian}). 

The unitary encoding the input displacement is complimented by control drives that entangle the qubit and oscillator via conditional displacements~\cite{diringer2023conditional} and qubit rotations (Fig.~\ref{Fig:Time-Independent}a). The entangling conditional displacements are applied before and after the unknown input is fed into the system, and the qubit is rotated by $\pi$ or $\pi/2$ pulses before, during, and after the input. Due to the qubit-state-dependent shift of the oscillator frequency by $-\chi$, these qubit rotations serve to make the oscillator sensitive to the input signal independent of the state of the qubit at the start of each round of input. Additionally, when combined with conditional displacements on the oscillator, the control and input scheme impart a geometric area enclosed by the oscillator trajectory onto the qubit, such that the phase of an unknown time-independent input signal can be extracted via a qubit measurement (see Appendix~\ref{sec:si:reservoir} for details of this unitary). In Appendix~\ref{sec:si:expressivity}, we show the ability of the set of unitaries implemented here to be able to approximate any scalar function of the input signal when the signal is time-independent. For all results presented, we implement our reservoir unitary with these control drives across all tasks, with 4 applications of the unitary interleaved with qubit and oscillator-parity measurements.

The binary classification task we perform here is: Two distributions of time-independent signals, completely characterized by the signal's in-phase ($I$) and quadrature ($Q$) components, are distributed along two separate ``arms of a spiral'' in the $I-Q$ plane (Fig.~\ref{Fig:Time-Independent}b). Given a displacement described by the points $I$ and $Q$ sampled from either signal distribution, one must figure out which distribution the signal came from. The maximum amplitude of the input signal distribution $\mathrm{max}(\vert \epsilon_{\mathrm{in}}\vert)$ (i.e. the points in the spiral arms furthest away from the origin in Fig.~\ref{Fig:Time-Independent}b) was chosen such that the amount of displacement of the oscillator state initialized in vacuum would result in a coherent state with $\bar{n} = 0.3$ photons per round of input ($\sim \SI{1}{\micro\second}$).  This input amplitude was needed in order to perform the classification with sufficient accuracy in a reasonable amount of shots. Our QRC solved the spiral classification task with $>$ 97\% accuracy at $10^3$ shots (Fig.~\ref{Fig:Time-Independent}b). This simple task has the feature that, if one feeds in the inputs directly into a linear layer, the classification accuracy would reach no more than $67\%$---just above the random guessing accuracy of $50\%$. As a point of comparison with non-linear digital reservoirs, we found that a 64-dimensional, two-layer digital reservoir was needed to achieve the same performance as our quantum reservoir for this task (see Appendix~\ref{sec:si:lesn:comparison} for details of this comparison).

To probe the role of quantum coherence in our reservoir, we performed the same classification task, but with reduced coherence time in the qubit during the reservoir execution (Fig.~\ref{Fig:Time-Independent}c). This was achieved by populating the lossy readout resonator with photons that send the qubit to the center of the Bloch-sphere when the readout resonator is traced out (see Appendix~\ref{sec:si:reservoir_characterization:t2}). With $T_2 \rightarrow 0$, we effectively removed all entanglement with the oscillator, and observed two things: a dramatic reduction in classification performance, and importantly, $T_2$ only began affecting the performance once it was on the order of the reservoir duration, after which the qubit is projected to a pure state. 

\subsection*{Classification of radio-frequency (RF) communication modulation schemes}\label{sec:results:rfml}
\addcontentsline{toc}{subsection}{\nameref{sec:results:rfml}}

\begin{figure*}[h]
\centering
    \includegraphics{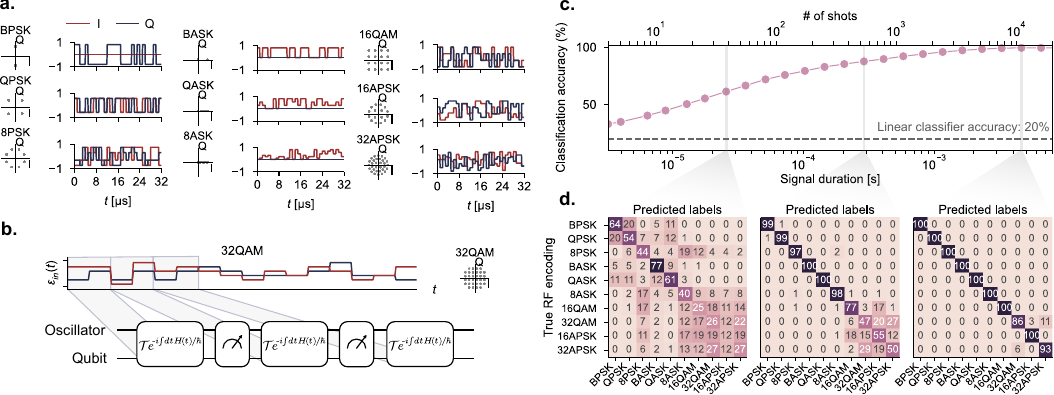}
    \caption{ 
    \textbf{ Classification of radio-frequency (RF) communication modulation protocols. }  
    \textbf{ (a) }  
        Description of the dataset for digital modulation schemes used in this experiment. In conventional digital modulation schemes, one encodes data in the amplitude and phase of the signal. The modulation schemes can be represented by a ``constellation diagram'' in $(I,Q)$ space (left), where point represents one of the possible choice of $(I,Q)$ values to encode a symbol, with example time traces (right).
    \textbf{(b)} 
        These signals are broken up and fed into our reservoir. 
    \textbf{(c)} 
        The performance of the reservoir as a function of the shots taken in real time (see text). The top row contains the corresponding duration of the radio frequency signal required. As the number of shots increases, the fluctuations in the measurement distribution reduces, resulting in a higher classification accuracy. For context, a classical linear classifier applied directly on the input  data achieves only $20\%$ accuracy, independent of the duration of the signal. The error bars of the accuracy curve have been omitted here due to the fact they are too small. 
    \textbf{(d)}
        Confusion matrix for the QRC at 32, 512, and $10^4$ shots, showing that the reservoir confuses only a few classes at the highest shots. 
    }\label{Fig:RFML}
\end{figure*}
Next, to highlight the ability to perform classification of higher dimensional data, we classified time-dependent radio-frequency (RF) signals. The microwave signals in this dataset encode digital information using one of 10 different digital modulation schemes, a standard benchmark task in RF machine learning~\cite{oshea2018over,JAGANNATH2021}. Digital modulation schemes encode binary information in discrete `symbols' encoding in sequential time-bins. For example, Binary Phase-Shift Keying (BPSK) encodes binary data in discrete phase jumps of a signal, such that a symbol $0$ ($1$) maps to a phase flip of $0$ ($\pi$). Other modulation schemes can encode more bits per symbol. BPSK and other encodings can be represented in a constellation diagram (Fig.~\ref{Fig:RFML}a), which denotes the potential $(I,Q)$ values a signal can take for each symbol. A given string of digital data can then be encoded in a time-domain signal by sequentially choosing points in the constellation diagram with a given symbol rate. For typical WiFi signals this is around 250 kHz per subchannel~\cite{RFML}.

For this task, we generated RF signals by encoding random digital strings into the 10 different modulation schemes with a fixed symbol rate of around 2 symbols per $\mu$s, or with a sampling rate of 2 MSps. The duration of these signals typically lasts much longer than the reset period of our system. Importantly, we did not repeat the same signal to artificially reduce the sampling noise associated with each input data, as this would not typically be applicable in a real-world setting. Instead, the measurement statistics were generated by sampling the signal in real time. Consequently, what we refer to as `shots' in a real-time task does not correspond to identical repetitions of the experiment, but instead, is the number of resets we performed while acquiring the signal, which changed from shot to shot. In effect, each different encoding scheme produces a unique ``fingerprint" distribution over measurement outcomes, and the goal of the linear layer is to separate these distributions with as high accuracy as possible. 

Figure~\ref{Fig:RFML}c shows the accuracy in classifying digitally modulated RF signals with increasing number of shots, compared with the performance of a linear classifier. We note that in less than a millisecond, or with less than 2000 symbols, the reservoir was able to classify which of the 10 classes a given signal belongs to with $>$ 90\% accuracy when using 8 qubit-oscillator measurements. A linear classifier can only achieve 20\% classification accuracy for this task, even with infinite symbols. The confusion matrix between the different classes at 32, 512, and $10^4$ shots is displayed in Fig~\ref{Fig:RFML}d, the latter two of which are nearly diagonal. 

\subsection*{Classification of filtered noise}\label{sec:results:noise_ml}
\addcontentsline{toc}{subsection}{\nameref{sec:results:noise_ml}}

\begin{figure*}[h]
    \centering
    \includegraphics[width=0.8\textwidth]{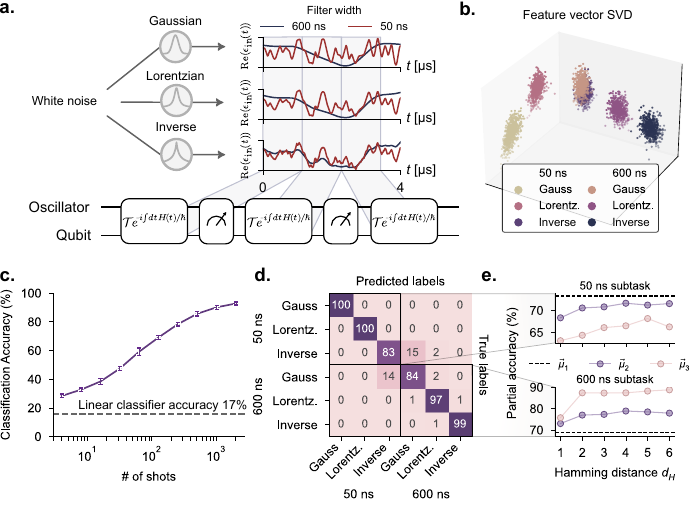}
    \caption{
    \textbf{Classification of filtered noise} 
    \textbf{(a)} 
        We classify various fast and slow noisy signals by applying a moving average on stochastic white-noise signals. Three different filters are used for the moving averages: a Gaussian filter, a Lorentzian, and an inverse power law. For each filter, we generate stochastic analog signals based on both a 50 ns filter width, and a 600 ns filter width, with the latter being on the order of the measurement rate. Example time traces are given for the real part of an example stochastic white noise signal passed through each of the filters. We divide up these stochastic signals and send to the QRC to then distinguish the noise source.
    \textbf{(b)}
        Visualization of the high-dimensional output feature space using Singular Value Decomposition (SVD). Each point corresponds to a different signal over 2000 shots taken in real time (see text). 
    \textbf{(c)}
        Classification accuracy as a function of the number of shots using third-order moments as the output feature. Our reservoir reaches 93\% accuracy in about 2000 shots, corresponding to about 10 ms of the signal received. 
    \textbf{(d)}
        Confusion matrix of the task taken at 2000 shots. 
    \textbf{(e)}
        Participation of the mean and the off-diagonal elements of the second- and third-order moments in the classification accuracy within the subtasks of classifying different noise sources with fixed filter width. We see that for signals with long coherence times, higher-order measurement correlations are important, while for fast signals, the mean dominates in the performance contribution. 
    }
    \label{Fig:Noise_ML}
\end{figure*}

Finally, to demonstrate the performance of our QRC on continuous-time data\footnote{The previous time-dependent task, RF-modulation-scheme classification, concerns discrete-time data.}, and with a task that requires both long-term and short-term memory in the quantum reservoir, we performed the following classification task: input data assumed to have come from a source of white noise is filtered using a moving-average filter having one of three filter shapes (Gaussian, Lorentzian and inverse-power-law), and one of two window widths ($50$ ns and $600$ ns), and the task is to identify both the filter shape and window width (Fig.~\ref{Fig:Noise_ML}a). The resultant dataset consisting of six classes of noisy signals was designed to probe the ability of our QRC to process high dimensional data with bandwidths larger than the cavity linewidth. Additionally, this task allowed us to probe the memory of our QRC and its ability to be sensitive to fluctuations in time, a key feature that enable temporal signal processing in QRCs~\cite{Sannia2024dissipationas,Pena2021}. The filter functions were normalized so that the photon-number distributions generated by the time-dependent displacements are identical up to the filter width. This normalization was applied to ensure that the task is not trivially solvable by just measuring the mean photon number (see Appendix~\ref{sec:si:tasks_si:noise_ml}).

Because all the signals used in this dataset are noise with zero mean, a linear classifier would do no better than random guessing. By contrast, Figure~\ref{Fig:Noise_ML}b visually shows (using Singular Value Decomposition (SVD) on the output feature space) that the quantum reservoir was able to peel apart the different noise distributions. On the task of classifying over six different sources of noise, we achieved 93\% accuracy (Fig.~\ref{Fig:Noise_ML}c) in only 2000 shots. As seen in the confusion matrix in Fig.~\ref{Fig:Noise_ML}d, the primary confusion at 2000 shots was distinguishing between the 50-ns inverse-power-law noise class and the 600-ns Gaussian noise class, as expected from the overlap in the SVD of the feature space. 

Finally, we compared the ability of our reservoir to understand long vs short correlations in input signals. For this, we deconstructed the 6-class classification task into two classification subtasks, where in each subtask, the QRC learned to distinguish noisy signals generated from among three different filter window types, but with fixed window widths. The two subtasks differ by the filter window width (see Fig.~\ref{Fig:Noise_ML}d and e). The class of signals with coherence length of 50 ns highlights the convenience of our input encoding scheme, i.e. feeding signals directly into the oscillator mode without the need to sample the signal discretely in time. \edit{Additionally, the ability for our quantum reservoir to distinguish between signals with correlation times on the order of 50 ns demonstrates the sensitivity to signals which vary on time-scales much faster than the measurement rate}. In contrast, classification of the class of signals with coherence lengths of $600$ ns requires correlations of the reservoir dynamics beyond that of the measurement rate. To highlight the advantage of our scheme, we simulated the performance of a reservoir with that of a recent gate-based protocol where the input was sampled discretely in time~\cite{yasuda2023quantum}. Our simulation results, in Appendix ~\ref{sec:si:simulation:continuous}, highlight the advantage of our protocol when the sampling rate of the input is slow, which can arise in experiment such as finite pulse durations and latency introduced by the FPGA classical comparison.

Figure~\ref{Fig:Noise_ML}e looks at the participation of the different moments $\mu_k$ of the measurements in the classification accuracy of the 50-ns subtask (top), and the 600-ns subtask (bottom). Here, the output features were constructed by the mean $\mu_1$, or the off-diagonal elements of the moments $\mu_2$ and $\mu_3$ as a function of the distance between measurements $d_H$, allowing us to probe the contribution of the moments as a function of the locality of the correlations. For the 50-ns subtask, we see that the most important contribution is the mean, with the second-order moment being the next-most important contribution, and the third-order moment being relatively unimportant. In stark contrast, the third-order moment is most important for the 600-ns subtask, surprisingly yielding nearly 90\% classification accuracy using non-local third-order correlations alone. \edit{} The ability to distinguish stochastic signals among the combined six classes demonstrates the ability of our reservoir to capture both slow and fast features of microwave signals.

\section*{Discussion}
\addcontentsline{toc}{section}{\nameref{sec:Discussion}}
\label{sec:Discussion}

In summary, we have experimentally realized an analog quantum reservoir computer (QRC) and demonstrated its ability to directly process microwave analog input signals without discretization, achieving high classification accuracy on three different tasks. Previous demonstrations of quantum reservoir computing have used multi-qubit, gate-based quantum reservoirs~\cite{yasuda2023quantum, Pfeffer_2022, PhysRevApplied.14.024065, Suzuki_2022, kubota2022quantum, mlika2023user, PhysRevX.13.041020}. In contrast, we perform machine learning directly on analog signals fed into a single oscillator coupled to a transmon qubit. The superconducting-circuits platform not only allows us to leverage projective non-demolition (QND) non-Gaussian measurements to generate correlated output features, but is also well-matched to process microwave signals that can generally be continuous in time. In addition to demonstrating accurate classification of microwave signals in our experiments, we also performed a direct comparison with a state-of-the-art discrete-time, gate-based QRC approach in simulation, and found that a continuous-time reservoir outperforms a discrete-time reservoir when the input signals contain temporal variations fast relative to the discretization time (see Appendix~\ref{sec:si:simulation:continuous}).

For any quantum neural network, including QRC approaches, a central concern is to what extent one can achieve high accuracy on a particular task without needing an impractical number of shots \cite{wright2019capacity}. Ref.~\cite{PhysRevX.13.041020} recently reported that certain functions---termed \textit{eigentasks}---can be constructed with low error from quantum reservoirs even when the number of shots is modest, giving evidence that for some tasks, sampling noise need not be overwhelming. In our experiments, we found that it was possible to achieve high accuracy for all the tasks we attempted while needing only $10^3$--$10^4$ shots (depending on the task). There is important future work to be done in exploring the trade-offs between reservoir size (e.g., number of oscillators or qubits), number of measurements $M$ between reservoir resets, feature-vector dimension (dependent both on $M$ and the choice of order of correlators to include), and number of shots required for both training and inference. 

With improved quantum hardware, we anticipate that it will be possible to carry out even more sophisticated tasks than what we have already demonstrated. Increasing the coherence time of the oscillator would enable us to perform many more measurements (the qubit's coherence time is, favorably, less important in our scheme because our protocol involves repeatedly projectively measuring the qubit). While we analytically showed in Appendix~\ref{sec:si:expressivity} the ability of our QRC to be able to approximate any scalar function of the input signal when the signal is time-independent, provided the number of measurements $M$ performed is large enough, there remains the open theoretical question of the expressiveness of the QRC when the input signal is time-dependent. \edit{Generalizing our approach to spatial in addition to temporal inputs, as was explored in Ref.~\cite{Spagnolo2022}, would likely support more sophisticated computations. In Appendix~\ref{sec:si:simulation:multi}, we explore such extensions in simulations and find a marked improvement in classification accuracy. }

It is an open question if QRC---using the type of reservoir we considered in this paper, or any other---can, when implemented with NISQ hardware, achieve a quantum computational advantage over the best classical machine learning approaches, just as it is unclear if any quantum-machine-learning method can \cite{cerezo2022challenges}. We did not investigate the potential for purely computational quantum advantage: our quantum reservoir is small enough to be easily classically simulable. However, our work opens up the possibility to experimentally achieve a different type of quantum advantage than a purely computational one. If one performs quantum processing on data obtained by a quantum sensor, there is the potential for an advantage that is a hybrid of being due to the advantage of quantum sensing and of quantum computing \cite{cerezo2022challenges}. Our work suggests the feasibility of concretely realizing this kind of hybrid quantum sensing-computational advantage, where the quantum sensor is a superconducting circuit that can detect classical microwave radiation with high quantum efficiency and low noise \cite{wang2021quantum, wang2022quantum, backes2021quantum, assouly2023quantum}. While the signals classified in this work originate at room temperature and are highly attenuated before reaching the device, our experiments have shown that it is possible to accurately classify signals using a superconducting circuit even when there are only a few photons of signal in the superconducting circuit within any single run. Combining this with a sensitive quantum detector could lead to quantum smart sensors---quantum versions of classical in-sensor processors \cite{zhou2020near}---that can reliably extract information from weak microwave signals in a way that exceeds the accuracy of any equivalent classical system.\\

\noindent
\textit{Note added}: During the final stages of our work, we became aware of a related effort, Ref.~\cite{hu2023overcoming}, and we coordinated to release our papers simultaneously. Ref.~\cite{hu2023overcoming} introduces a protocol for quantum reservoir computing with temporal data. Similar with theirs, our approach also uses mid-circuit measurements. We experimentally realized our reservoir with an analog quantum system, in contrast to their implementation, which was with a discrete-time, gate-based quantum system.

\section*{Data and code availability}
All data generated and code used in this work is
available at: \url{https://doi.org/10.5281/zenodo.10432778}

\section*{Author contributions}

A.S. designed and carried out the hardware experiments and performed the data analysis. S.P. performed the numerical simulations of the quantum system and helped to optimize the experimental protocol with early contribution from J.K.. V.K. performed the numerical simulations of the classicalized quantum system, and performed the comparisons with classical machine-learning methods. V.F. oversaw the design and creation of the superconducting device by S.R. and others. A.S. and V.F. set up the cryogenic and microwave apparatus. S.R. calibrated the superconducting device with A.S. and V.F.. Y.C. and X.W. performed the theoretical analysis of the expressivity in Appendix~\ref{sec:si:expressivity}. T.O., L.G.W. and P.L.M. conceived the project, and T.O. and J.K. performed initial numerical simulations to validate the concept. A.S., S.P. and P.L.M. wrote the manuscript with input from all authors. P.L.M. supervised the project.

\section*{Acknowledgements}
The authors would like to thank Hakan T{ü}reci, Shyam Shankar, Saeed A. Khan, Haohai Shi, William P. Banner, Shi-Yuan Ma, and Maxwell Anderson for helpful discussions and comments. The authors would also like to thank Bradley Cole, Clayton Larson, Britton Plourde, Eric Yelton, and Luojia Zhang for the fabrication of the transmon and on-chip resonator, Chris Wang for the design of the transmon, the on-chip resonator and the 3D superconducting cavity (using pyEPR~\cite{pyEPR}), and Nord Quantique for the fabrication of the 3D superconducting cavity. We gratefully acknowledge MIT Lincoln Laboratory for supplying the Josephson traveling-wave parametric amplifier (TWPA) used in our experiments. This paper is based upon work supported by the Air Force Office of Scientific Research under award number FA9550-22-1-0203. We gratefully acknowledge a DURIP award with AFOSR award number FA9550-22-1-0080 for equipment used in this work. The authors wish to thank NTT Research for their financial and technical support. PLM acknowledges membership in the CIFAR Quantum Information Science Program as an Azrieli Global Scholar. Y.C. and X.W. were supported by the Air Force Office of Scientific Research under Grant No. FA9550211005, NSF CCF-1942837 (CAREER), and a Sloan Research Fellowship.

\clearpage

\section*{Methods}\label{sec:methods}
\addcontentsline{toc}{section}{\nameref{sec:methods}}
\setcounter{figure}{0}% Restart figure numbering
\renewcommand{\thefigure}{M\arabic{figure}}% Figure counter representation
\renewcommand{\theHfigure}{M\arabic{figure}}% Hyperref figure hyperlink hook

\subsection*{Reservoir unitary}\label{sec:methods:reservoir_unitary}
\addcontentsline{toc}{subsection}{\nameref{sec:methods:reservoir_unitary}}

To design a good reservoir computer capable of performing machine learning on a variety of tasks, one needs to implement control drives that can efficiently capture important information of the input and perform a non-trivial and non-linear map to output features. Here, our reservoir is composed of alternating unitaries and measurements. The design of the former is motivated to harness the quantum properties of the dynamical system to generate entanglement and the design of the latter to generate non-linear operations on the state of our reservoir via measurement back-action. Here we summarize the control drives and measurements we use and their effect on the reservoir dynamics, both in the context of time-dependent and time-independent signals. 

For time-independent signals, the unitary implemented in our reservoir (see Fig.~\ref{Fig:Time-Independent}b) can be approximated by the following set of unitaries (see Appendix~\ref{sec:si:reservoir})
\begin{align}
    U_1 &= X_{\pi/2} \label{eq:methods:reservoir_decomp_u1}\\ 
    U_2 &= D(\alpha) \vert g \rangle \langle e \vert + D(-\alpha) \vert e \rangle \langle g \vert && \text{CNOD} \label{eq:methods:reservoir_decomp_u2}\\ 
    U_3 &= D(\beta) \vert g \rangle \langle g \vert + \vert e \rangle \langle e \vert && \text{Input} \label{eq:methods:reservoir_decomp_u3}\\
    U_4 &= X_{\pi} \label{eq:methods:reservoir_decomp_u4}\\ 
    U_5 &= U_3 = D(\beta) \vert g \rangle \langle g \vert + \vert e \rangle \langle e \vert  && \text{Input} \label{eq:methods:reservoir_decomp_u5}\\
    U_6 &= D(-\alpha) \vert g \rangle \langle e \vert + D(\alpha) \vert e \rangle \langle g \vert && \text{CNOD} \label{eq:methods:reservoir_decomp_u6}\\
    U_7 &= Y_{\pi/2}. \label{eq:methods:reservoir_decomp_u7}
\end{align}
This combination of unitaries encloses a loop in the oscillator's phase space. The area of this closed loop, which depends on the phase of the unknown displacement $\beta$, imparts a geometric phase onto the qubit. In this work, we perform this unitary directly after a qubit measurement without reset. The action of the combined unitary on the qubit prepared in the ground or excited state, and for an arbitrary oscillator state, is
\begin{align}
    U \vert g \rangle  &= U_7U_6U_5U_4U_3U_2U_1 \vert g \rangle \\
      &= \frac{1}{\sqrt{2}} D(\beta) [ i\sin(A - \pi/4) \vert g \rangle  + \cos(A - \pi/4) \vert e \rangle ]\otimes \vert \mathrm{cavity} \rangle \\ 
      U \vert e \rangle  &= \frac{1}{\sqrt{2}} D(\beta) [ i\cos(A - \pi/4) \vert g \rangle  + \sin(A - \pi/4) \vert e \rangle ]\otimes \vert \mathrm{cavity} \rangle
\end{align}
where $A = 2 \vert \alpha \vert \vert \beta \vert \sin (\delta) = i(\alpha \beta^* - \alpha^* \beta)$ is the geometric phase enclosed by the oscillator trajectory, and dependent on the phase difference $\delta$ between a known displacement $\alpha$, and the unknown displacement $\beta$. The probability of measuring the qubit in the excited state given it started out in the ground state, $P_{e \vert g}$, and the probability of measuring excited given the qubit start in the excited state, $P_{e \vert e}$ are given by 
\begin{equation}
    P_{e \vert g} = \cos(A - \pi/4)^2 \qquad \qquad P_{e \vert e} = \sin(A - \pi/4)^2
\end{equation}
The equation relates the qubit probability to the phase of the input displacement, which is otherwise challenging to extract in a setup with only qubit measurements. 

For general time-dependent signals, the closed loop formed by Eqs.~\ref{eq:methods:reservoir_decomp_u1}-\ref{eq:methods:reservoir_decomp_u7} is broken, and the system is entangled before the measurement. While this can be hard to study analytically in the general case, we take a look at a special case of time-dependent signals, namely those of Fig.~\ref{Fig:RFML}. Here, the signal is time-dependent up to half the duration, so that the signal is effectively two time-independent signals combined. As a result, Eqs.~\ref{eq:methods:reservoir_decomp_u3}~and~\ref{eq:methods:reservoir_decomp_u5}~are no longer equal, but each still a time-independent displacement, and thus, the effects of the cross-Kerr, as discussed in Appendix~\ref{sec:si:reservoir}, do not hinder the interpretation of the effective gate-based model. For such input signals, the state of the system just before measurement is
\begin{equation}
    \vert \psi \rangle  = \frac{1}{2} [e^{iA_i}D(\beta_i) + e^{-iA_j}D(\beta_j)] \vert g, \mathrm{cavity} \rangle + \frac{1}{2} [e^{iA_i}D(\beta_i) - e^{-iA_j}D(\beta_j)] \vert e, \mathrm{cavity} \rangle,
    \label{eq:methods:final_state}
\end{equation}
where $\beta_i$ is the displacement just before the qubit flip (corresponding to Eq.~\ref{eq:methods:reservoir_decomp_u3} for this time-dependent set of tasks), and $\beta_j$ is the displacement after (Eq.~\ref{eq:methods:reservoir_decomp_u5}). $A_i = \alpha \beta_i$ is the phase acquired after two non-orthogonal displacements. When $\beta_i = \beta_j$ we recover the dynamics for time-independent signals.

\subsection*{Repeated measurements}\label{sec:methods:repeated_measurements}
\addcontentsline{toc}{subsection}{\nameref{sec:methods:repeated_measurements}}

The unitaries described above are followed by a qubit measurement, then a parity measurement. For time-independent signals, the qubit and oscillator are disentangled at the end of the unitary, and the effect of the unitary on the oscillator is just a displacement. Thus we can ignore any affects of the qubit measurement on the oscillator. The state of the oscillator after $M$ repeated measurements and $M$ time-independent displacements can be effectively described as
\begin{equation}
    \vert \mathrm{cavity} \rangle = \ldots P_{p_4} D(\beta)P_{p_3} D(\beta)P_{p_2} D(\beta)P_{p_1} D(\beta) \vert 0 \rangle,
    \label{eq:si:repeated_cavity}
\end{equation}
where $P_{x_n}$ is the projector of the $n$th parity measurement $\Pi$ with measurement outcomes $x_n = \{+,-\}$. In Appendix~\ref{sec:si:expressivity}, we show that by sampling the parity measurements alone combined with the linear layer, we can realize (but not limited to) the following vector space of funtions: 
\begin{equation}
    \mathcal{H}_\mathrm{parity} := \left\{ c_0 + c_1 e^{-2 \vert \beta \vert^2} + c_2 \left( e^{-2\vert \beta \vert^2} \right)^2 + \cdots + c_k \left( e^{-2 \vert \beta \vert ^2} \right)^k: c_0, c_1, \dots, c_k \in \mathbb{R} \right\}.
\end{equation}

\subsection*{Output feature encoding \& the linear layer}\label{sec:methods:feature_vector}
\addcontentsline{toc}{subsection}{\nameref{sec:methods:feature_vector}}

In reservoir computing, the outputs of a reservoir, called feature vectors, are sent to a trained linear layer. Here, we briefly outline the motivation and construction of the feature vectors and the training algorithms used in this manuscript. 

In general, sampling over all possible measurement trajectory outcomes and generating a probability distribution contains all the information one can extract from a quantum system. However, not all the information plays an equal role for finite samples. Thus, for our work here, we use a physically motivated output feature vector that efficiently captures the relevant information for a linear layer. The output feature vectors for our reservoir are generated from computed correlations of measurement outcomes. The $p$-th order correlations are characterized by the $p$-th central moment $\mu_p$ of the underlying distribution of measurement trajectories. The elements of $\mu_p$ are 
\begin{equation}
    (\mu_p)_{ijkl\ldots} = \frac{1}{N_{\mathrm{shots}}}\sum_n^{\mathrm{N_{shots}}} (x_{ni} - \langle x_{i} \rangle ) (x_{nj} - \langle x_{j} \rangle ) (x_{nk} - \langle x_{k} \rangle )(x_{nl} - \langle x_{l} \rangle ) \ldots,
    \label{eq:methods:central_moments}
\end{equation}
where $x_{in}$ is the $n$th repeated measurement outcome of observable $x_i$ for a total of $N_{\mathrm{shots}}$ repetitions, and $\langle \ldots \rangle$ is the expectation value taken over repetitions. For the results presented in the main text, we use only up to third-order correlations. Additionally, due to the finite memory present in our reservoir, we only keep correlations between nearest, next-nearest, and next-next-nearest measurements. See Appendix~\ref{sec:si:ML:output_feature_encoding} for details and motivation behind this choice. 

For machine learning with reservoir computing, the only component of the reservoir that is trained is a linear layer applied to the above feature vectors. The linear layer is an $R \times C$ matrix $W_{\mathrm{train}}$ and applied to the $R$-dimensional feature vector $x$, and biased with a $C$-dimensional vector $v_{\mathrm{train}}$:
\begin{equation}
    y = W_{\mathrm{train}}x + v_{\mathrm{train}}.
\end{equation}
Here $C$ is equal to the number of classes in the data set. The largest elements of $y$ corresponds to the class that the reservoir predicts the given input data point $x$ belongs to. To train the weight matrix $W_{\mathrm{train}}$, we either use a pseudo-inverse method to minimize the mean squared error (MSE) between $W_{\mathrm{train}}x$ and $y$, or backpropagation to minimize the MSE after a softmax function. Both methods are described in more detail in Appendix~\ref{sec:si:ML:linear_layer}. In the main manuscript, we present results for whichever performed the best.

\newgeometry{text={\dimexpr8.5in-40mm,\dimexpr11in-50mm}}
% reversemp, top={26mm}, headheight={5.5pt}, headsep={5.6mm}, marginparsep=5mm, marginparwidth=12mm, footskip=10mm}

%%\input sn-article.bbl
% \nocite{*}
% \bibliographystyle{} % Override
% \bibliography{%
\putbib[%
 bib/general,%
 bib/qrc_ibm,%
 bib/qrc_theory,%
 bib/qrc_general,%
 bib/qrc_hakan,%
 bib/rc_general%
]
% }
\restoregeometry
\end{bibunit}
\pagebreak

%%%%%%%%%%%%%%%%%%%%%%%%%%%%%%%%%%%%%
%%%%%%%%%%%%%%%%%%%%%%%%%%%%%%%%%%%%%

\pagebreak
\begin{appendices}
\begin{bibunit}

\setcounter{figure}{0}% Restart figure numbering
\setcounter{table}{0}% Restart table numbering
\renewcommand{\thefigure}{S\arabic{figure}}% Figure counter representation
\renewcommand{\theHfigure}{S\arabic{figure}}% Hyperref figure hyperlink hook
\renewcommand{\thetable}{S\arabic{table}}% Table counter representation

\startcontents[sections] % local ToC
{
\hypersetup{linkcolor=black}
% \renewcommand{\contentsname}{Appendix Contents}
% \tableofcontents
\section*{Appendix Contents}
\printcontents[sections]{}{0}{}
}

\vspace{\baselineskip}
{\centering \noindent\rule{0.5\normaltextwidth}{0.4pt}\par}
\vspace{\baselineskip}

\newpage

%%%%%%%%%%%%%%%%%%%%%%%%%%%%%%%%%%%%%%%%%%%%%%%%%%%%%%%%%%%%%%%%%%%%%%%%%%%%%%%%%%%%
%%%%%%%%%%%%%%%%%%%%%%%%%%%%%%%%%%%%%%%%%%%%%%%%%%%%%%%%%%%%%%%%%%%%%%%%%%%%%%%%%%%%
%%%%%%%%%%%%%%%%%%%%%%%%%%%%%%%%%%%%%%%%%%%%%%%%%%%%%%%%%%%%%%%%%%%%%%%%%%%%%%%%%%%%
\section{Experimental setup}\label{sec:si:experimental_setup}

\begin{figure*}[h]
    \centering
    \includegraphics[width=0.88\textwidth]{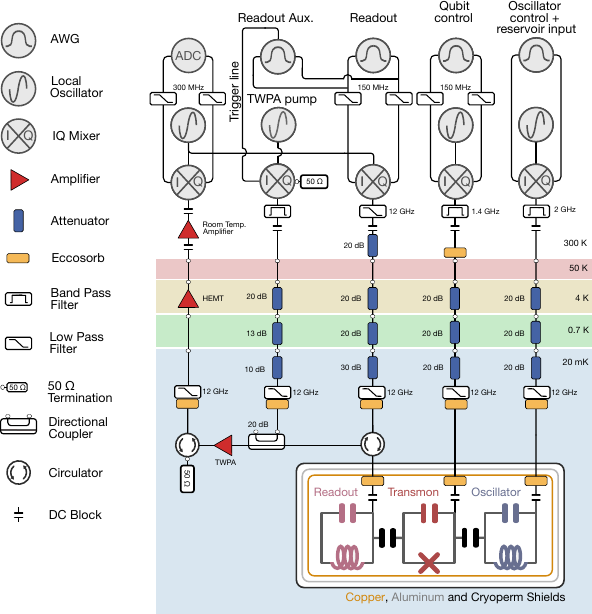}
    \caption{\textbf{Wiring diagram.} Experimental setup for control hardware, cable routing, and shielding for our device.}
    \label{fig:si:wiring_diagram}
\end{figure*}

\begin{figure*}[h]
    \centering
    \includegraphics[width=0.85\textwidth]{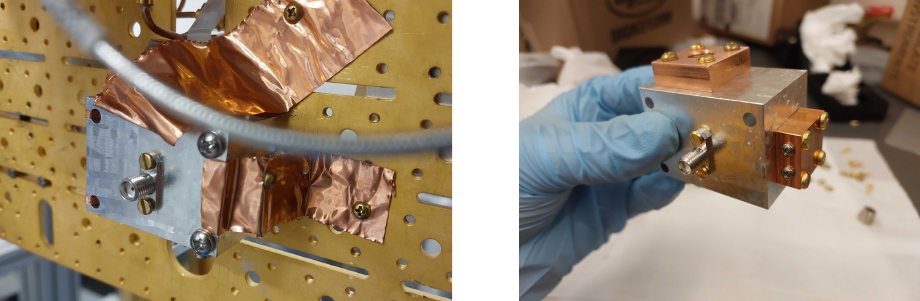}
    \caption{\textbf{ Photo of device} 
        The device consists of a on-chip transmon and a co-planar waveguide readout resonator mounted inside a high-purity Aluminum cavity. The package is mounted to a gold-plated copper breadboard at the mixing-chamber plate of a dilution refrigerator. 
    }
    \label{fig:si:device_photo}
\end{figure*}

The device used in this paper consists of an oscillator, a 3D stub post cavity made from high-purity 4N Aluminum treated with an acid etch, and a transmon qubit. The transmon, made of Niobium, is fabricated on a resistive silicon chip, along with an on-chip readout resonator also made of Niobium. The single chip hosting the transmon and the readout resonator is mounted in the 3D cavity package using copper clamps. The cavity and the copper clamp contain copper films for thermalization directly to gold-plated copper breadboard at the mixing chamber plate of the dilution refrigerator (Fig.~\ref{fig:si:device_photo}). The device is shielded with Copper coated with Berkeley Black, and two types of magnetic shields: Aluminum, and Cryoperm (Fig.~\ref{fig:si:wiring_diagram}). The cavity pin is set such that the oscillator mode is undercoupled to the transmission line by a factor of 40. While this reduces the transmission of photons incident on our device by a factor of 40, it keeps the oscillator state thermalized to the fridge rather than the transmission line.

The control pulses for the qubit and the storage are synthesized using Zurich Instruments (ZI) HDAWG, which have a baseband bandwidth of 1 GHz. These are upconverted using Rohde \& Schwarz SGS100A, which are signal generators with built-in IQ mixers. These built-in mixers are used for all frequency conversions with the exception of the readout. The readout pulses are synthesized and digitized using ZI UHFQA, and are up-converted and down-converted using Marki mixers (MMIQ-0416LSM-2), with a split LO from a single SGS100A. Readout signals are first amplified with a traveling-wave Josephson Amplifier (TWPA), which is a quantum-limited amplifier. The TWPA typically requires large pump tones, so we gate it with a trigger line from the readout AWG which combines with the CW pump tone in an IQ mixer (as a makeshift fast switch). The readout signals are then further amplified with a High-electron mobility transistor (HEMT) ampliflier at the 4K stage, and again amplified with a room temperature amplifier (ZVA-1W-103+ from Mini-Circuits) and filtered. The digitizer on the ZI UHFQA converts to the analog response to a digital signal and integrates it to produce a binary outcome depending on the qubit state. 

For the experiments that intentionally suppress the qubit $T_2$ via resonator induced dephasing via pumping of the readout resonator, we use an additional ZI HDAWG channel that combines with the AWG of the ZI UHFQUA. This was mostly a choice out of convenience, as the AWG of the ZI UHFQUA has limitations that made characterizations tricky.

%%%%%%%%%%%%%%%%%%%%%%%%%%%%%%%%%%%%%%%%%%%%%%%%%%%%%%%%%%%%%%%%%%%%%%%%%%%%%%%%%%%%
%%%%%%%%%%%%%%%%%%%%%%%%%%%%%%%%%%%%%%%%%%%%%%%%%%%%%%%%%%%%%%%%%%%%%%%%%%%%%%%%%%%%
%%%%%%%%%%%%%%%%%%%%%%%%%%%%%%%%%%%%%%%%%%%%%%%%%%%%%%%%%%%%%%%%%%%%%%%%%%%%%%%%%%%%
\section{System Hamiltonian \& Reservoir description}\label{sec:si:reservoir}

\subsection*{Hamiltonian description}\label{sec:si:reservoir:hamiltonian}
\addcontentsline{toc}{subsection}{\nameref{sec:si:reservoir:hamiltonian}}

Our transmon-cavity system is well approximated by the Hamiltonian~\cite{Blais_2021}:
\begin{equation}
    H/\hbar = \omega_q q^{\dagger}q  + \omega_a a^{\dagger}a - \chi q^{\dagger}q a^{\dagger}a - \chi' q^{\dagger}q a^{\dagger 2} a^2 - K_q q^{\dagger 2} q^2  - K a^{\dagger 2} a^2 + \edit{\Xi}(t) (q + q^{\dagger}) +  \edit{\xi}(t)(a + a^{\dagger}),
    \label{eq:si:full_hamiltonian}
\end{equation}
where $a$ is the annihilation operator for the oscillator mode, and $q$ is the annihilation operator for the qubit mode, $\omega_a$ and $\omega_q$ are the frequencies of the oscillator and qubit mode respectively, $\chi$ and $\chi'$ are the dispersive shift and the oscillator state-dependent dispersive shift respectively, $K$ and $K_q$ are the self-Kerr of the oscillator and the transmon anharmonicity respectively. The values for these parameters, as well as values for decay rates, are listed in Table~\ref{table:si:system_parameters}. \edit{The last two terms describe the qubit and oscillator drives in the lab frame. The lab-frame drives are related to the rotating-frame drives in Eq.~\ref{eq:Hamiltonian} via $\Xi(t) = \Omega(t)e^{i\omega_q t} + \mathrm{H.c.}$ and $\xi(t) = \epsilon(t)e^{i\omega_a t} + \mathrm{H.c.}$}. For the design of our drives, we ignore the self-Kerr of the oscillator as well as the higher-order cross-Kerr. We note that these are indeed present, but for the purposes of a quantum reservoir, only add to the complexity of the dynamics. Finally, moving to the rotating frame of the transmon and oscillator mode and truncating to the first two levels of the transmon, we arrive at the Hamiltonian in Eq.~\ref{eq:Hamiltonian}. 

\begin{table}[h]
\centering
\begin{tabular}{ c|c|c||c } 
 Parameter & Mode(s) & Symbol & Value \\ 
 \hline
 \hline
 Frequency & Transmon g-e & $\omega_{q}$ & 2$\pi \times$ 7.136 GHz \\ 
           & Oscillator & $\omega_{a}$ & 2$\pi \times$ 6.024 GHz \\ 
           & Readout & $\omega_{r}$ & 2$\pi \times$ 8.888 GHz \\ 

  Self-Kerr & Transmon g-e & $K_q$ & $2\pi \times$ 315 MHz  \\ 
           & Oscillator & $K$ & $2\pi \times$ 6 kHz \\ 
  Cross-Kerr & Transmon-Oscillator & $\chi$ & 2$\pi \times$ 2.415 MHz \\ 
           & Transmon-Readout & $\chi_{qr}$ & 2$\pi \times$ 1 MHz \\ 
  Second-order Cross-Kerr & Transmon-Oscillator & $\chi'$ & 2$\pi \times$ 19 kHz \\ 
  \hline
   Relaxation time & Transmon g-e & $T_1$ & 30 $\mu$s \\ 
           & Oscillator & $T_1^{a}$ & 100 $\mu$s \\ 
 Dephasing time & Transmon g-e & $T_2$ & 25 $\mu$s \\ 
 Thermal population & Transmon g-e & $\bar{n}_{\mathrm{eq}}^q$ & 3\% \\ 
           & Oscillator & $\bar{n}_{\mathrm{eq}}^a$ & $<$ 0.2\% \\ 
\end{tabular}
 \caption{%
 \textbf{System parameters and dissipation rates.} 
    System parameters were measured using various spectroscopic and time-domain techniques following methods in Ref.~\cite{chou2018teleported}.
 }
  \label{table:si:system_parameters}
\end{table}

%%%%%%%%%%%%%%%%%%%%%%%%%%%%%%%%%%%%%%%%%%%%%%%%%%%%%%%%%%%%%%%%%%%%%%%%%%%%%%%%%%%%
%%%%%%%%%%%%%%%%%%%%%%%%%%%%%%%%%%%%%%%%%%%%%%%%%%%%%%%%%%%%%%%%%%%%%%%%%%%%%%%%%%%%
%%%%%%%%%%%%%%%%%%%%%%%%%%%%%%%%%%%%%%%%%%%%%%%%%%%%%%%%%%%%%%%%%%%%%%%%%%%%%%%%%%%%
\subsection*{Reservoir description for time-independent signals}\label{sec:si:reservoir:time_independent}
\addcontentsline{toc}{subsection}{\nameref{sec:si:reservoir:time_independent}}

\begin{figure*}[h]
    \centering
    \includegraphics[width=0.7\textwidth]{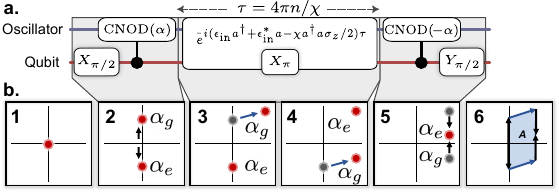}
    \caption{
    \textbf{Geometric phase unitary to sense phase of unknown displacement}
    \textbf{(a.)}
        Decomposition of the unitary used throughout the reservoir for time-independent tasks.
    \textbf{(b.)}
        Schematic representation of the dynamics of the oscillator state under the reservoir drives with time-independent input, highlighting a unitary which implements a geometric phase unitary. (1) At the start of the protocol, the qubit is in the ground state, with the oscillator at vacuum. (2) An initial $X_{\pi/2}$ pulse brings the qubit to the equator. The $\mathrm{CNOD}(\alpha)$ unitary conditions the state of the oscillator based on the qubit. For the first reservoir, this is a coherent state. (3) and (4) For time independent inputs, the effective action can be described by a single displacement on the oscillator mode. In this experiment, we operate the displacement at a frequency which, to first order, causes a displacement only on the state conditioned on the ground state of the qubit. A qubit $\pi$ pulses switches the state in between the two displacements. (5) The final conditional displacement brings the two conditioned states back onto each other. This effectively disentangles the qubit from the oscillator mode. (6) The effective geometrical area enclosed $A$, which is a function of the input, is imparted onto the qubit.
    }
    \label{fig:si:reservoir}
\end{figure*}

The advantage of the reservoir computing paradigm is the flexibility in the choice of dynamics. However, simple design principles, motivated by the physics of the system, can go a long way in engineering a reservoir with high expressive capacity on many tasks. In this section, we provide full details and motivations for the unitaries and measurements in this work, followed by sections outlining characterizations of the device in order to realize the intended dynamics. 

The reservoir drives consists of two categories of dynamics: the unitaries and the measurements. In what follows, we will first provide analysis of the dynamics for time-independent input (e.g. the signals in Fig.~\ref{Fig:Time-Independent}). As we will see, the unitary component of the dynamics implemented in this work strives to implement a $\cos^2$ nonlinearity on the raw input, whereas the measurements generate non-classical features in the state and quantum correlations in the measurement trajectories via measurement backaction. 

 While measuring the quadratures of some unknown signal is easy with a typical homodyne setup, performing the same measurement of a displacement on an oscillator using only qubit measurements can be non-trivial. Of course, when designing a reservoir, one does not strive to implement the identity, but it is a good starting point -- the unitary is thus implemented to approximate the identity. It consists of the input signal data, which is sandwiched on either side by fast conditional displacement gates implemented with CNOD~\cite{diringer2023conditional} and qubit rotation gates. The broad-overview of the decomposed unitary is given in terms of gates in Fig.~\ref{fig:si:reservoir}, along with a schematic portrayal of the phase-space trajectory of the oscillator mode initialized in vacuum subject to a time-independent drive. 

We begin with an idealized gate-based version decomposition of our reservoir for time-independent input on resonance with the oscillator conditioned on the qubit being in the ground state. The sequence of gates the reservoir unitary approximates:
\begin{align}
    U_1 &= X_{\pi/2} \label{eq:si:reservoir_decomp_u1}\\ 
    U_2 &= D(\alpha) \vert g \rangle \langle e \vert + D(-\alpha) \vert e \rangle \langle g \vert && \text{CNOD} \label{eq:si:reservoir_decomp_u2}\\ 
    U_3 &= D(\beta) \vert g \rangle \langle g \vert + \vert e \rangle \langle e \vert && \text{Input} \label{eq:si:reservoir_decomp_u3}\\
    U_4 &= X_{\pi} \label{eq:si:reservoir_decomp_u4}\\ 
    U_5 &= U_3 = D(\beta) \vert g \rangle \langle g \vert + \vert e \rangle \langle e \vert  && \text{Input} \label{eq:si:reservoir_decomp_u5}\\
    U_6 &= D(-\alpha) \vert g \rangle \langle e \vert + D(\alpha) \vert e \rangle \langle g \vert && \text{CNOD} \label{eq:si:reservoir_decomp_u6}\\
    U_7 &= Y_{\pi/2} \label{eq:si:reservoir_decomp_u7}
\end{align}
Ignoring the very first unitary, after applying the sequence of unitaries $U_2$ through $U_7$, we arrive at unitary
\begin{equation}
    U_7U_6U_5U_4U_3U_2 = \frac{i}{\sqrt{2}}e^{(\alpha \beta^* - \alpha^*\beta)}D(\beta) (\vert g \rangle \langle g \vert - \vert e \rangle \langle g \vert  ) - \frac{i}{\sqrt{2}}e^{(-\alpha \beta^* + \alpha^*\beta)}D(\beta) (\vert g \rangle \langle e \vert + \vert e \rangle \langle e \vert ) 
    \label{eq:reservoir_unitary}
\end{equation}
Let 
\begin{equation}
    \vert \psi \rangle = [e^{-i\phi/2}\cos(\theta/2)\vert g \rangle + e^{i\phi/2}\sin(\theta/2) \vert e \rangle ]\otimes \vert \mathrm{cavity} \rangle 
\end{equation}
be some arbitrary initialized state. Then for $\theta = \pi/2$, we have
\begin{equation}
    U_7U_6U_5U_4U_3U_2 \vert \psi \rangle = \frac{1}{\sqrt{2}} D(\beta) [ i\sin(A - \phi/2) \vert g \rangle  + \cos(A - \phi/2) \vert e \rangle ]\otimes \vert \mathrm{cavity} \rangle,
\end{equation}
where $A = 2 \vert \alpha \vert \vert \beta \vert \sin (\delta) = i(\alpha \beta^* - \alpha^* \beta)$ is the geometric phase enclosed by the oscillator trajectory which is dependent on the phase difference $\delta$ between the known displacement $D(\alpha$), and the unknown displacement $D(\beta)$ (Fig~\ref{fig:si:reservoir}b). Thus, for the proper qubit state before the application of $U_2\ldots U_7$, we are able to extract information about the phase of the displacement. We also note that the qubit and the oscillator are disentangled after the unitary, and that the effect of the unitary on the oscillator mode is a simple displacement. 

Finally, pre-pending $U_1$ (Eq.~\ref{eq:si:reservoir_decomp_u1}) to the string of unitaries guarantees that we initialize our qubit state with $\theta = \pi/2$ when following a qubit measurement, independent of that measurement outcome. It also guarantees $\phi = \pi/2$ or $3\pi/2$ depending on the measurement outcome.  The probability of measuring the qubit in the excited state conditioned on preparing it $e$ vs $g$ after the entire sequence is then:
\begin{equation}
    P_{e \vert g} = \cos(A - \pi/4)^2 \qquad \qquad P_{e \vert e} = \sin(A - \pi/4)^2
    \label{eq:si:reservoir_unitary_pe}
\end{equation}
Thus, with this sequence of unitaries, we are able to extract the phase of some unknown displacement (relative to some known displacement $\alpha$) by simply measuring the qubit. While for the first run of the reservoir, the qubit will start in the ground state (up to thermal noise), after performing a parity measurement, the qubit state will depend on the previous measurement outcome. See Fig.~\ref{fig:si:phase_sensing} for an experimental implementation of the above results. 

In principle, Eq.~\ref{eq:si:reservoir_unitary_pe} enables us to perform the identity operation on the input $x,y$ points followed by a $\cos^2$ kernel. Without loss of generality, we take $\arg(\alpha) = 0$, then $ i(\alpha \beta^* - \alpha^* \beta) = \mathrm{Im}(\beta) = \beta_x$. Alternating between $\arg(\alpha) = 0$ and $\arg(\alpha) = \pi/2$ allows us to extract $\cos^2(\vec{\beta})$ with two runs of the reservoir. 

Whereas all gates besides the input (Eqs.~\ref{eq:si:reservoir_decomp_u3}~and~\ref{eq:si:reservoir_decomp_u5}) are fast and therefore insensitive to the cross-Kerr interaction, the primary deviation from the gate description occurs for the input, which can be very long. This input displacement is conditioned on the qubit being in the ground state. Therefore, in the rotating from of the qubit-oscillator system, the branch of the oscillator state conditioned on the qubit being in the excited state will rotate at a frequency $\chi$, which in general will break the geometric phase construction that works for time-independent tasks. Therefore, we limit the exposure time of the reservoir to the input signal to be an integer multiple of $4\pi/\chi$, so that the oscillator state conditioned on the qubit being in the excited state will return to the same point. 

The unitary described in Eqs.~\ref{eq:si:reservoir_decomp_u1}-\ref{eq:si:reservoir_decomp_u7}~is followed by a qubit measurement, then a parity measurement $\Pi$~\cite{Heeres_2015,wang2021noise} with projectors $P_{\pm}$, where
\begin{equation}
    \Pi = (-1)^{a^{\dagger}a} \qquad \qquad P_{\pm} = \frac12 (1 \pm \Pi)
    \label{eq:si:parity}
\end{equation}
As mentioned above, the effect of the unitary on the oscillator state for time-independent signals is simply a displacement of the input data $D(\beta)$, independent of the qubit measurement outcome. For the following discussion, we will ignore the qubit dynamics, since the qubit and the oscillator are disentangled at the end of the unitary. In effect, the state of the oscillator can be described by a series of alternating displacements and parity measurements:
\begin{equation}
    \vert \mathrm{cavity} \rangle = \ldots P_{p_4} D(\beta)P_{p_3} D(\beta)P_{p_2} D(\beta)P_{p_1} D(\beta) \vert 0 \rangle,
    \label{eq:si:repeated_cavity}
\end{equation}
where $P_{p_n}$ is the projector of the $n$th parity measurement with outcomes $p_n = \{+,-\}$. For $k$ runs of the reservoir, we can reorder terms and add pairs of canceling displacements $D(-\beta)D(\beta)$ to rewrite the above as
\begin{equation}
    \vert \mathrm{cavity} \rangle = \left( \prod_n^k P_{p_n}^{n\beta} \right) D(k\beta) \vert 0 \rangle.
    \label{eq:si:contextuality}
\end{equation}
Equation~\ref{eq:si:contextuality} describes a series of projective measurements after preparing a displaced vaccum state. The projectors and their associated measurements are
\begin{equation}
    P_{\pm}^{\alpha} = D(\alpha)P_{\pm}D(-\alpha) \qquad \qquad \Pi^{\alpha} = D(\alpha)\Pi D(-\alpha)
    \label{eq:si:effective_measurements}
\end{equation}
The measurements $\Pi^{\alpha}$ describe parity measurements in displaced frame at $\alpha$. Incidentally, the expectation value of this operator are proportional to the Wigner function at $\alpha$~\cite{Royer1977}. However, importantly, Eq.~\ref{eq:si:contextuality} \textit{does not} describe performing Wigner tomography of the state $D(k\beta)\vert 0 \rangle = \vert k\beta \rangle$ at points given by $\beta, 2\beta, 3\beta, \ldots$, as the effective measurements $\Pi^{\alpha}$ do not commute for different values of $\alpha$. Instead, in general $[\Pi^{\alpha},\Pi^{\gamma}] \ne 0$. Therefore, in this light, our reservoir construction can be seen to leverage non-commuting measurements and quantum contexuality to generate conditional and correlated probabilities over measurement trajectories. 

%%%%%%%%%%%%%%%%%%%%%%%%%%%%%%%%%%%%%%%%%%%%%%%%%%%%%%%%%%%%%%%%%%%%%%%%%%%%%%%%%%%%
%%%%%%%%%%%%%%%%%%%%%%%%%%%%%%%%%%%%%%%%%%%%%%%%%%%%%%%%%%%%%%%%%%%%%%%%%%%%%%%%%%%%
%%%%%%%%%%%%%%%%%%%%%%%%%%%%%%%%%%%%%%%%%%%%%%%%%%%%%%%%%%%%%%%%%%%%%%%%%%%%%%%%%%%%

\subsection*{Reservoir description for slow varying time-dependent signals }\label{sec:si:reservoir:time_dependent}
\addcontentsline{toc}{subsection}{\nameref{sec:si:reservoir:time_dependent}}

\begin{figure*}[h]
    \centering
    \includegraphics[width=0.7\textwidth]{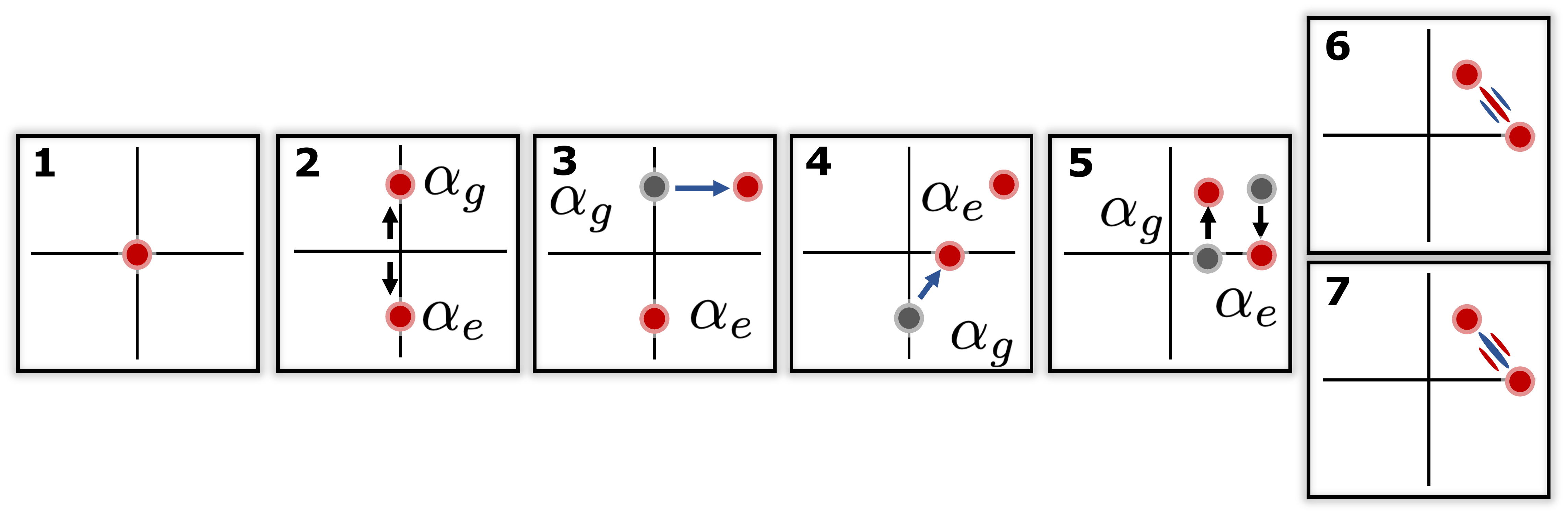}
    \caption{\textbf{Schematic description of the dynamics of the oscillator, starting in vacuum, for a slow time-varying input} In such a scenario (as is the case for the task of classifying radio modulation schemes), the signal causes a displacement largely conditioned on the ground state. Generally, the value of the displacement is different before and after the $\pi$ pulse in between (3) and (4). Unlike the regime for time-independent signals, there is no effective area enclosed in phase space, which leaves the qubit entangled with the oscillator. (6) and (7) describe the state of the oscillator after the qubit measurement. The resulting state of the oscillator is a cat state, where the parity of the cat dependent on whether the outcome of the qubit measurement is ground or excited.}
    \label{fig:si:reservoir_rfml}
\end{figure*}

For generic, time-dependent signals, like those classified in Figs.~\ref{Fig:RFML} and~\ref{Fig:Noise_ML} in the main text, the geometric unitary described by Eqs.~\ref{eq:si:reservoir_decomp_u1}-\ref{eq:si:reservoir_decomp_u7} does not in general hold, as the symmetry between panels 3 and 4 in Fig.~\ref{fig:si:reservoir} is broken. Additionally, the approximation that the input is displacement conditioned on the qubit in the ground state (Eq.~\ref{eq:si:reservoir_decomp_u3} and~\ref{eq:si:reservoir_decomp_u5}) will not hold for high bandwidth signals, like those in Fig.~\ref{Fig:Noise_ML} in the main text. For high-bandwidth signals, the input will also have some contribution in displacing oscillator conditioned on the qubit being in the excited state, which can lead to complex dynamics in the oscillator. While for generic signals, this can be hard to describe, here we prove a treatment of our reservoir construction for slowly-varying, time-dependent signals, like those in Fig.~\ref{Fig:RFML} of the main text.

We can follow most of the derivation from the scenario of time independent signals in Appendix~\ref{sec:si:reservoir:time_independent}, to describe the dynamics of the QRC for the task of classifying radio frequency modulation schemes. Along with the assumptions in the previous section, we make the slow-varying input approximation, such that the displacement on the oscillator of the reservoir is still effectively conditioned on the ground state of the qubit. The displacement on the oscillator depends on the value of the symbol encoded for the given modulation scheme. Since, in general, the symbol is different before and after the qubit $\pi$ pulse: the direction of the displacement in the oscillator will be different. Given the timescales of the input signal involved, this essentially corresponds to a displacement on the oscillator conditioned on the ground state of the qubit. When the two displacements are different in magnitude and direction, the qubit remains entangled with the oscillator at the end of the reservoir unitary. The state of the system just before the measurements is (step (5) of Fig~\ref{fig:si:reservoir_rfml}:

\begin{equation}
    \vert \psi \rangle = \frac{1}{\sqrt{2}}(e^{-iA_i}D(\beta_i)\vert g, \mathrm{cavity} \rangle + e^{iA_j}D(\beta_j)\vert e, \mathrm{cavity} \rangle),
\end{equation}

where $\beta_i$ is the displacement before the qubit flip, and $\beta_j$ is the displacement after. $A_i = 2\text{Im}\left[\alpha \beta_i^*\right]$ is the phase acquired after two non-orthogonal displacements. When $\beta_i = \beta_j$, we recover the dynamics for time independent signals. It is straightforward to show that the qubit will be disentangled from the oscillator and that the area $A_i$, corresponding to the geometrical phase form the area enclosed in phase space will be present as a relative phase difference between the ground and excited state. After a $Y_{\pi/2}$ gate, we have the following state in our system:

\begin{equation}
    \vert \psi \rangle  = \frac{1}{2} [e^{iA_i}D(\beta_i) + e^{-iA_j}D(\beta_j)] \vert g, \mathrm{cavity} \rangle + \frac{1}{2} [e^{iA_i}D(\beta_i) - e^{-iA_j}D(\beta_j)] \vert e, \mathrm{cavity} \rangle,
\end{equation}

One can think of this as a cat state in the cavity, with a parity determined by the qubit state. This is schematically shown in (6) and (7) in Fig~\ref{fig:si:reservoir_rfml}. In the limit of very different displacements, the probability of the qubit measurement is the same for both ground and excited states. The goal of this task can be thought of as discriminating probability distribution functions over the $(I,Q)$ plane. Fig~\ref{Fig:RFML} (a) represents the so-called ``constellation " diagram of the modulation schemes considered in this work. Each scheme can take discrete values in $(I,Q)$ space, with even equal probability (we construct the dataset of radio signals encoding random binary strings). Our lack of knowledge of the exact displacement on the oscillator can be mathematically expressed as a density matrix. This is the most apparent in the state of the oscillator after the initial qubit measurement,

\begin{equation}
    \rho_{\mathrm{cavity}}' =  \sum_{\beta_i \in P} p_i D(\beta_i)\rho_{\mathrm{cavity}} ^\dag D(\beta_i) ,
\end{equation}

where $\rho_{\mathrm{cavity}}'$ is the density matrix representation of the cavity right after the qubit measurement and $\rho_{\mathrm{cavity}}$ describes the initial density matrix before the application of the protocol. The set $P$ describes the distribution of possible displacements which can be received from the input. $p_i$ is probability for receiving the symbol corresponding to a displacement $\beta_i$
% , and $p_{ij}$ is the conditional probability of displacement $\beta_j$, given $\beta_i$. 
For the task considered in this work, these distributions are uniform, with no contributions from conditional probabilities. However, this description of the reservoir motivates the potential for the QRC to distinguish signals with complex correlations in the symbols of the message encoded. 

%%%%%%%%%%%%%%%%%%%%%%%%%%%%%%%%%%%%%%%%%%%%%%%%%%%%%%%%%%%%%%%%%%%%%%%%%%%%%%%%%%%%
%%%%%%%%%%%%%%%%%%%%%%%%%%%%%%%%%%%%%%%%%%%%%%%%%%%%%%%%%%%%%%%%%%%%%%%%%%%%%%%%%%%%
%%%%%%%%%%%%%%%%%%%%%%%%%%%%%%%%%%%%%%%%%%%%%%%%%%%%%%%%%%%%%%%%%%%%%%%%%%%%%%%%%%%%

\section{Quantum reservoir characterization}\label{sec:si:reservoir_characterization}

%%%%%%%%%%%%%%%%%%%%%%%%%%%%%%%%%%%%%%%%%%%%%%%%%%%%%%%%%%%%%%%%%%%%%%%%%%%%%%%%%%%%
%%%%%%%%%%%%%%%%%%%%%%%%%%%%%%%%%%%%%%%%%%%%%%%%%%%%%%%%%%%%%%%%%%%%%%%%%%%%%%%%%%%%
%%%%%%%%%%%%%%%%%%%%%%%%%%%%%%%%%%%%%%%%%%%%%%%%%%%%%%%%%%%%%%%%%%%%%%%%%%%%%%%%%%%%

\subsection*{CNOD}\label{sec:si:reservoir_characterization:cnod}
\addcontentsline{toc}{subsection}{\nameref{sec:si:reservoir_characterization:cnod}}

\begin{figure*}[h]
    \centering
    \includegraphics[width=0.7\textwidth]{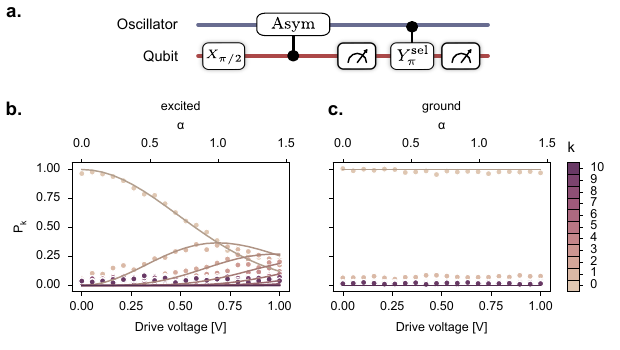}
    \caption{
    \textbf{Characterization of conditional displacements using an Anti-symmetric pulse from Ref.~\cite{diringer2023conditional}.}
    \textbf{(a.)}
    Pulse control schematic of characterization of conditional displacement. Here, we make use of number-splitting spectroscopy to characterize the state of the cavity after perform a conditional displacement on the state $\vert +\rangle \vert \mathrm{vacuum} \rangle$ conditioned on the qubit state. By post-selecting on the qubit state, we can evaluate the effectiveness of the conditional displacement. 
    \textbf{(b.)}
    Number splitting spectroscopy conditioned on measuring the qubit in the excited state. A single parameter fit is used to capture the behavior of the state of the cavity as a function of amplitude. From the good agreement, we conclude that conditioned on measuring the qubit in the excited state, the cavity is displaced.  
    \textbf{(c.)}
    Number splitting spectroscopy conditioned on measuring the qubit in the ground state. We see very limited change in the cavity state when measuring the qubit in the ground state.
    }
    \label{fig:si:asymmetric_pulse}
\end{figure*}

Here, we provide the calibration of the CNOD unitary~\cite{diringer2023conditional}, one of the components of our reservoir unitary (Fig.~\ref{fig:si:reservoir}). The CNOD protocol implements the following unitary
\begin{equation}
    \mathrm{CNOD}(\alpha) = D(\alpha)\vert g \rangle \langle e \vert + D(-\alpha)\vert e \rangle \langle g \vert.
    \label{eq:si:cnod}
\end{equation}
The protocol is implemented with two `Anti-symmetric pulses' sandwiching a qubit pi-pulse. In the frequency domain, the pulse is composed of two gaussian envelopes offset such that there is a zero-crossing at the qubit ground state frequency, and that the spectrum is anti-symmetric around this point (see Ref.~\cite{diringer2023conditional}). The Anti-symmetric pulse is a conditional displacement, conditioned on the qubit being in the excited state.  The motivation for using CNOD instead of a single tone displacement on resonance with the stark-shifted qubit frequency is that it enables the ability to perform conditional displacements at time scales much smaller than $2\pi/\chi$. 

Figure~\ref{fig:si:asymmetric_pulse}a displays the protocol for characterizing the anti-symmetric pulse. First, the qubit is unconditionally brought to the equator of the bloch sphere, with a wide-band $X_{\pi/2}$ pulse. After this, the anti-symmetric pulse acts on the cavity, followed by an qubit measurement, collapsing the cavity state to either $D(\alpha)\vert 0 \rangle$ or $\vert 0 \rangle$. After collapsing the state, we perform a number-splitting spectroscopy on the cavity. This is performed with a conditional $Y_{\pi}$, conditioned on the $k$th cavity Fock state~\cite{Gambetta2006,Schuster2007} followed by a second qubit measurement. By post-selecting on the first qubit measurement outcome, we can characterize the cavity state for each branch. Figure~\ref{fig:si:asymmetric_pulse}b and c show the number-splitting spectroscopy for the cavity state conditioned on the qubit being in the ground state vs excited, as a function of pulse amplitude. These curves are fitted with a single parameter scaling parameter that defines the relationship between pulse amplitude voltage and the amount of displacement $\alpha$.

%%%%%%%%%%%%%%%%%%%%%%%%%%%%%%%%%%%%%%%%%%%%%%%%%%%%%%%%%%%%%%%%%%%%%%%%%%%%%%%%%%%%
%%%%%%%%%%%%%%%%%%%%%%%%%%%%%%%%%%%%%%%%%%%%%%%%%%%%%%%%%%%%%%%%%%%%%%%%%%%%%%%%%%%%
%%%%%%%%%%%%%%%%%%%%%%%%%%%%%%%%%%%%%%%%%%%%%%%%%%%%%%%%%%%%%%%%%%%%%%%%%%%%%%%%%%%%
\subsection*{Reservoir unitary characterization}\label{sec:si:reservoir_characterization:characterization}
\addcontentsline{toc}{subsection}{\nameref{sec:si:reservoir_characterization:characterization}}

\begin{figure*}[h]
    \centering
    \includegraphics[width=0.70\textwidth]{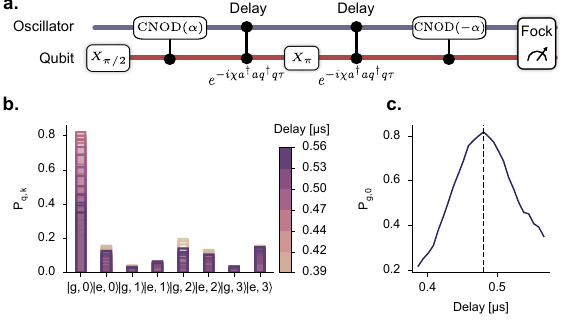}
    \caption{
    \textbf{Characterization of input signal duration }
    \textbf{(a.)}
        Diagram of control sequence to calibrate length of input signal duration toward implementing the geometric phase unitary in Eqs.~\ref{eq:si:reservoir_decomp_u1}-\ref{eq:si:reservoir_decomp_u7}. Here, a double conditional displacement is performed after sending the qubit to the equator of the Bloch sphere. After this, a variable delay is added before undoing that displacement. Finally, the qubit and the Fock distribution of the cavity is sampled using methods from Refs.~\cite{PhysRevX.10.021060,deng2023heisenberglimited}. 
    \textbf{(b.)}
        Overlapping histograms showing Fock distribution of cavity state conditioned on qubit state as a function of delay. 
    \textbf{(c.)}
        Cavity state overlap with the vacuum state as a function of delay. At a particular value of the delay, the two displacements interfere and cancel each other out. 
    }
    \label{fig:si:displace_undisplace}
\end{figure*}

With our rotation gates and CNOD's calibrated, we describe in this section the calibration of signal drives toward the implementation of Eqs.~\ref{eq:si:reservoir_decomp_u1}-\ref{eq:si:reservoir_decomp_u7}. We begin with a calibrating the duration of time our reservoir is exposed to the input signal. As discussed in Appendix~\ref{sec:si:reservoir}, calibrating this delay is crucial for a faithful implementation of the geometric phase detection unitary introduced in this work. While it may seem that this restriction in the input signal duration is contrived in a real-world setting where the signal is unknown, this restriction would be implemented via a fast switch that exposes our device to the unknown signal periodicially. 

Figure~\ref{fig:si:displace_undisplace}a schematically describes the experimental protocol for calculating the delay between the two CNOD pulses. Here, we effectively try to undo a double conditional displacement via second double conditional displacement. Due to the dispersive shift, after the first conditional displacement, the state of the cavity conditioned on the excited state of the qubit will start rotating with respect to the state of the cavity conditioned on the ground state. After a period of $2\pi/\chi$, the this will return to the same position as the start. Undoing the displacement at this point in time will send the cavity state to vacuum. Figure~\ref{fig:si:displace_undisplace}b shows the Fock distribution of the cavity as a function of the waiting time, and Fig.~\ref{fig:si:displace_undisplace}c shows the cavity state overlap with the vacuum state as a function of the waiting time. 

\begin{figure*}[h]
    \centering
    \includegraphics[width=0.85\textwidth]{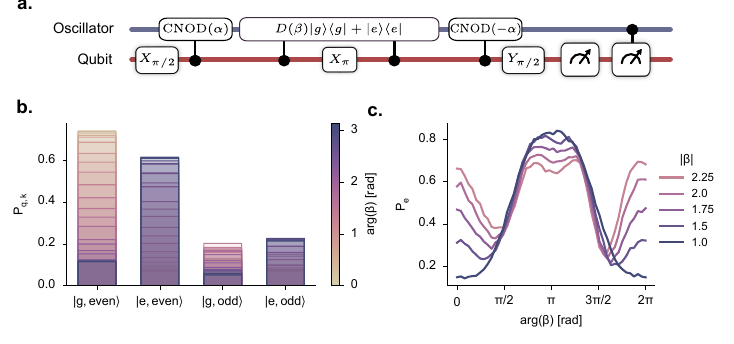}
    \caption{
    \textbf{Full geometric phase unitary calibration. }
    \textbf{(a.)}
        Pulse control protocol for the unitary defined by Eqs~\ref{eq:si:reservoir_decomp_u1}-\ref{eq:si:reservoir_decomp_u7}. This protocol is identical to that of Fig.~\ref{fig:si:reservoir}a.
    \textbf{(b.)}
        Overlapping histograms showing the probability over outcomes of measuring the qubit and oscillator-parity as a function of the phase of the input signal with $\vert \beta \vert  = 0.25$. As the phase is varied, the probability of the qubit being excited increases. 
    \textbf{(c.)}
        The probability of measuring the qubit excited as the phase of the input displacement $\beta$ is varied, plotted for different values of $\vert \beta \vert$
    }
    \label{fig:si:phase_sensing}
\end{figure*}

Next, we implement the full unitary given by Eqs.~\ref{eq:si:reservoir_decomp_u1}-\ref{eq:si:reservoir_decomp_u7}, where the section corresponding to the input data displacement (Eqs.~\ref{eq:si:reservoir_decomp_u3} and~\ref{eq:si:reservoir_decomp_u5}) is given by the duration found in the results above. For this calibration, we implement the full unitary given by the diagram in Fig.~\ref{fig:si:reservoir}a by varying the angle of the input displacement and looking at the dependence. 

Figure~\ref{fig:si:phase_sensing}a shows the schematic overview of the calibration procedure. The geometric phase unitary is parameterized by a long displacement, whose angle we sweep. After the unitary we perform a qubit measurement, followed by a parity measurement. This calibration experiment is essentially identical to the time-independent reservoir computing experiments in terms of the control protocol. Here, instead of sending data from different distributions for the system to classify, we only vary the phase and amplitude of some input displacement to get the phase dependence we want. 

Figure~\ref{fig:si:phase_sensing}b shows the distribution of measurement outcomes from measuring the qubit and the oscillator parity after the unitary is applied with $\alpha = 1$ and $\beta = 0.25$. As the angle of the input is swept, the qubit probability of the qubit being in found in the ground state shifts to being found in the excited state. This is more evident in Fig.\ref{fig:si:phase_sensing}c where we plot the probability of measuring the qubit in the excited state $P_e$ as a function of the phase of $\beta$ for different amplitudes of $\beta$. In comparison we find great qualitative agreement with the expected result $P_e = \cos(2 \vert \alpha \vert \vert \beta \vert \cos(\delta) + \pi/4)^2$, where $\delta = \arg(\alpha) - \arg(\beta)$ (see Eq.\ref{eq:si:reservoir_unitary_pe}), though we find an extra reduction in the dynamic range in $P_e$ for increasing $\beta$ due to qubit overheating.

For our quantum reservoir tasks, we choose $\alpha$ to be quite small, near 0.2. The effect of this is a severe reduction in the dynamic range of $P_e$, but one that is easily distinguishable at 1000 shots. For all of our tasks, this was the mininum number of shots needed to get 100\%. Keeping $\vert \alpha\vert $ small allows for a greater sensitivity in $\vert \beta \vert$ without worrying about qubit overheating. 

%%%%%%%%%%%%%%%%%%%%%%%%%%%%%%%%%%%%%%%%%%%%%%%%%%%%%%%%%%%%%%%%%%%%%%%%%%%%%%%%%%%%
%%%%%%%%%%%%%%%%%%%%%%%%%%%%%%%%%%%%%%%%%%%%%%%%%%%%%%%%%%%%%%%%%%%%%%%%%%%%%%%%%%%%
%%%%%%%%%%%%%%%%%%%%%%%%%%%%%%%%%%%%%%%%%%%%%%%%%%%%%%%%%%%%%%%%%%%%%%%%%%%%%%%%%%%%
\subsection*{Qubit \& parity measurements}\label{sec:si:reservoir_characterization:measurements}
\addcontentsline{toc}{subsection}{\nameref{sec:si:reservoir_characterization:measurements}}

\begin{figure*}[h]
    \centering
    \includegraphics[width=0.45\textwidth]{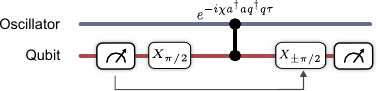}
    \caption{
    \textbf{Qubit and oscillator-parity measurements in the quantum reservoir computer. }
    For our reservoir construction, the state of the qubit is not generally known before a parity measurement. We apply feedback to change the parity measurement based on the preceding qubit measurement to faithfully capture the oscillator parity. 
    }
    \label{fig:si:qubit_parity_measurement}
\end{figure*}

The qubit and parity measurements performed in this work are the standard pulse schemes used in many previous works, with one change. The typical procedure of measuring the parity of a cavity state is similar to a Ramsey experiment (and perhaps more closer still to a `qubit-revival' experiment~\cite{chou2018teleported}), and importantly requires knowledge of the state of the qubit before the measurement is performed. In a quantum reservoir setting where measurement trajectories can be unknown, measuring the parity of the cavity is not straight-forward without post-selection or feedback. Here, since we perform a qubit measurement just before the parity measurement, we apply simple feedback that conditions the parity unitary on the measurement outcome of the preceding qubit measurement. The condition is such that the parity measurement outcome is now independent of the preceding measurement outcome. This reduces the order of correlations required to gain the same information: attaining the parity of the cavity only requires information about the parity measurement, whereas previously, second-order correlations between the qubit and parity measurement was required. A further refinement to reduce trivial correlations in the measurement history would reset the qubit after the oscillator parity, however, due to limitations in the FPGA software, this was not implemented. 

%%%%%%%%%%%%%%%%%%%%%%%%%%%%%%%%%%%%%%%%%%%%%%%%%%%%%%%%%%%%%%%%%%%%%%%%%%%%%%%%%%%%
%%%%%%%%%%%%%%%%%%%%%%%%%%%%%%%%%%%%%%%%%%%%%%%%%%%%%%%%%%%%%%%%%%%%%%%%%%%%%%%%%%%%
%%%%%%%%%%%%%%%%%%%%%%%%%%%%%%%%%%%%%%%%%%%%%%%%%%%%%%%%%%%%%%%%%%%%%%%%%%%%%%%%%%%%
\subsection*{Tuning $T_2$ via resonator-induced dephasing}\label{sec:si:reservoir_characterization:t2}
\addcontentsline{toc}{subsection}{\nameref{sec:si:reservoir_characterization:t2}}

\begin{figure*}[h]
    \centering
    \includegraphics[width=0.7\textwidth]{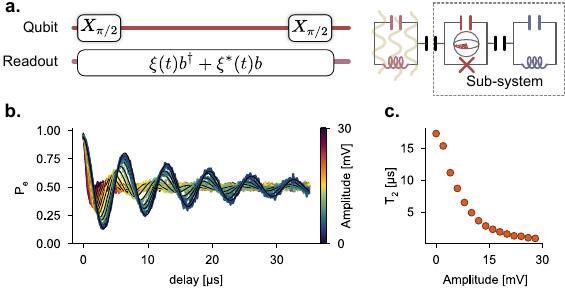}
    \caption{
    \textbf{ Resonator-induced dephasing via pumping on the readout resonator  }
    \textbf{(a.)}
        Protocol for characterizing the effect of a readout pump on qubit decoherence: we perform a typically Ramsey experiment while populating the readout resonator to measure the $T_2$. 
    \textbf{(b.)}
        Ramsey curves as a function of readout pump amplitude. These curves are fit using Eq.~\ref{eq:si:t2} to produce estimates of the qubit coherence. 
    \textbf{(c.)}
        Extracted $T_2$ values for each of the curves in part (b)
    }
    \label{fig:si:resonator_induced_dephasing}
\end{figure*}

Here we describe the experiment to reduce the qubit coherence time by pumping the readout resonator with photons during our reservoir experiments (see Fig.~\ref{Fig:Time-Independent}d). The calibration of this experiment involves performing a standard Ramsey $T_2$ experiment, modified with a pump on the readout resonator (Fig.~\ref{fig:si:resonator_induced_dephasing}a). Once populated, the resonator photons induce a dispersive shift, which sends the qubit to the center of the Bloch sphere once the readout resonator is traced out. In principle, this interaction is coherent, and the qubit should see a revival. However, due to the leaky nature of the readout resonator by design, a coherent revival is not observed. As remarked at the end of Appendix~\ref{sec:si:experimental_setup}, this experiment required an auxiliary AWG line. Figure~\ref{fig:si:wiring_diagram} denotes this as the `Readout Auxiliary' line.

Figure~\ref{fig:si:resonator_induced_dephasing}b shows the results of the Ramsey calibration with the readout pump on, for varying pump powers. We see a steady decrease in the qubit coherence time as the pump amplitude is increased as expected. The curves are fit to the equation
\begin{equation}
    P_e =  \cos (2\pi \delta t)  e^{-t/T_2},
    \label{eq:si:t2}
\end{equation}
where $\delta$ is an intentional detuning. Here, a Gaussian pulse was used as the readout pump. We expect that due to the construction of the reservoir, a flattop pulse may be more detrimental to the classification performance, since the Gaussian pulse has little amplitude during the CNOD unitaries shown in Fig.~\ref{fig:si:reservoir}a. Finally, we note that the maximum $T_2$ shown in Fig.~\ref{fig:si:resonator_induced_dephasing} differs from the value quoted in Table~\ref{table:si:system_parameters}. After preliminary calibration data corresponding to those in Fig.~\ref{fig:si:resonator_induced_dephasing}, the experiments in Fig.~\ref{Fig:Time-Independent}c were performed, after which the qubit $T_2$ was suddenly lowered. However, all experiments presented in this manuscript, with the exception of Fig.~\ref{fig:si:resonator_induced_dephasing}, were performed where the qubit $T_2$ matched that of Table~\ref{table:si:system_parameters}. Given the conclusion that the qubit $T_2$ does not impact classification accuracies until it approaches the time between measurements, we decided to include the higher quality data presented in Fig.~\ref{fig:si:resonator_induced_dephasing}, rather than the preliminary data used to calibrate the results in Fig.~\ref{Fig:Time-Independent}c.

%%%%%%%%%%%%%%%%%%%%%%%%%%%%%%%%%%%%%%%%%%%%%%%%%%%%%%%%%%%%%%%%%%%%%%%%%%%%%%%%%%%%
%%%%%%%%%%%%%%%%%%%%%%%%%%%%%%%%%%%%%%%%%%%%%%%%%%%%%%%%%%%%%%%%%%%%%%%%%%%%%%%%%%%%
%%%%%%%%%%%%%%%%%%%%%%%%%%%%%%%%%%%%%%%%%%%%%%%%%%%%%%%%%%%%%%%%%%%%%%%%%%%%%%%%%%%%
\section{Machine learning with the quantum reservoir}\label{sec:si:ML}
\subsection*{Output feature encoding}\label{sec:si:ML:output_feature_encoding}
\addcontentsline{toc}{subsection}{\nameref{sec:si:ML:output_feature_encoding}}

\begin{figure*}[h]
    \centering
    \includegraphics[width=0.5\textwidth]{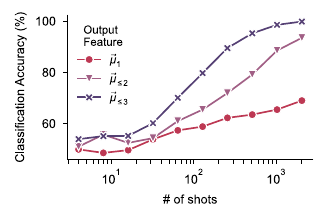}
    \caption{
    \textbf{ Comparison of feature vectors in spiral classification performance  } Here, classification accuracy on the spiral task is considered for different output encodings. Particularly, we compare included higher and higher correlations. $\vec{\mu}_{\leq p }$ describes a feature vector containing all central moments up to and including the $p$-th central moment (see text).    }
    \label{fig:si:spiral_accuracy_correlations}
\end{figure*}

In this work, we use measurement correlations as the output feature vectors from which the trained linear layer of our reservoir performs the classification. In this section, we provide details in how these were constructed from measurement results, as well as motivations and comparisons with other output encodings. As described in the main text, measurements of our reservoir involve two measurements following every data input: a qubit measurement and a parity measurement. The qubit measurement, which follows just after the input unitary, either extracts information about the input displacement (if the signal is time-independent), or performs some nontrivial back-action on the oscillator state (see Fig.~\ref{fig:si:reservoir_rfml}). The parity measurement, which follows the qubit measurement, will simply measure the parity of the cavity state post-qubit measurement, and collapse the oscillator state to either even or odd Fock states. It is worth pointing out that measurements of the parity are done with an entangling unitary starting with a known qubit state and then performing a regular qubit measurement (see Appendix~\ref{sec:si:reservoir_characterization:measurements} for details).

In this manuscript, qubit measurements are performed using standard dispersive readout, which we review here, since the process involves a number of nonlinear steps (for a thorough review, see Ref.~\cite{Blais_2021}). Each measurement outcome is the result of integrating a response signal from the readout resonator, and is defined by a single point on the $I-Q$ plane. For sufficiently strong coupling between the readout resonator and the qubit compared with the resonator linewidth, the set of all possible integrated IQ points will form two (or more) localized and well-seperated blobs, indicating projective measurement with single-shot fidelity. These two (or more) blobs correspond to different states of the transmon, and single-shot fidelity refers to the ability to discern the state of the qubit using only one readout pulse. With knowledge of the location of these blobs, and which state they correspond to, we perform a threshold the measurement result to either `0' or `1', indicating the qubit ground state or excited state respectively. 

From a string of binary measurement outcomes, or bitstring, we form our feature vectors by first calculating the $p$-th central moment $\mu_p$, defined as
\begin{equation}
    (\mu_p)_{ijkl\ldots} = \frac{1}{N_{\mathrm{shots}}}\sum_n^{\mathrm{N_{shots}}} (x_{ni} - \langle x_{i} \rangle ) (x_{nj} - \langle x_{j} \rangle ) (x_{nk} - \langle x_{k} \rangle )(x_{nl} - \langle x_{l} \rangle ) \ldots,
    \label{eq:si:central_moments}
\end{equation}
where the number of indices of $\mu_p$ is equal to $p$. Here $x_{ni}$ is the $n$th repeated measurement result of observable $x_i$. In our setting, $i$ labels the $i$-th measurement in a sequence of correlated measurements before the system is reset. The expectation value $\langle x_i \rangle$ is taken over the shots $N_{\mathrm{shots}}$ -- counting the number of system resets and repetitions. Faithful estimates of these moments typically require on the order of 1000 shots for the results presented in this manuscript.

\begin{figure*}[h]
    \centering
    \includegraphics[width=0.6\textwidth]{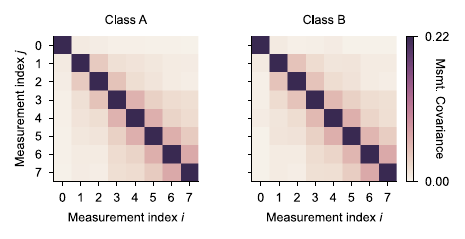}
    \caption{
    \textbf{ Second-order central moment (covariance) of quantum reservoir output over spiral dataset. }  These correlation matrices were generated from calculating the covariance over measurement outcomes in a reservoir run, then averaged over the entire dataset.   }
    \label{fig:si:spiral_correlations2}
\end{figure*}

The central moments of Eq.~\ref{eq:si:central_moments}~are used in the construction of the output feature vector for the linear layer to perform the classification task. Specifically, the feature vector is generated by appending successively more and more central moments. We denote these appended feature vectors as $\vec{\mu}_{\leq p}$ for feature vectors containing up to $p$ central moments, e.g. 

\begin{align*}
    \vec{\mu}_{\leq 2} &= [\vec{\mu}_1,\mu_2] \\  
    & = [ \langle x_0 \rangle, \langle x_1 \rangle,\langle x_2 \rangle,\ldots, \langle x_0 x_1 \rangle - \langle x_0 \rangle \langle x_1 \rangle, \langle x_0 x_2 \rangle - \langle x_0 \rangle \langle x_2 \rangle,\ldots, \langle x_1 x_2 \rangle - \langle x_1 \rangle \langle x_2 \rangle,\ldots]
\end{align*}is a feature vector constructed from appending the flattened covariance to the mean. The first-order moment here is a vector to denote we take the mean over repetitions of different measurements, whereas the covariance is a matrix and thus is not denoted as a vector. Additionally, we only take the independent degrees of freedom of the symmetric covariance matrix equivalent to discarding one of the following redundant elements $\langle x_i x_j \rangle$ and $\langle x_j x_i \rangle$ for some integers $i,j$. In general, for arbitrary moments, the number of independent components for $M$ measurements is ${M+p-1}\choose{p}$, where $p$ is the order of the moment. For up to third-order central moments of $M = 8$, this gives a total output feature dimension of $\mathrm{dim}(\vec{\mu}_{\leq 3}) = 8 + 36 + 120 = 164 $. This output dimension for the results presented in the main text is further reduced as discussed in the following paragraphs below.

\begin{figure*}[h]
    \centering
    \includegraphics[width=\textwidth]{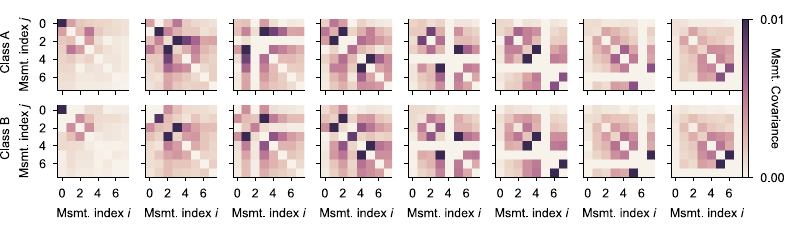}
    \caption{
    \textbf{ Third-order central moment of quantum reservoir output over spiral dataset.}     
    These third-order central moments are plotted as an array of 2D matrices, such that the $i$-th column corresponds to the 2D matrix $\mu_{ijk}$. The two rows corresponds to the two different classes of signals. In the third order, one can begin to see differences between the two classes by eye. 
    }
    \label{fig:si:spiral_correlations3}
\end{figure*}

Figure~\ref{fig:si:spiral_accuracy_correlations} contains classification results on the spiral dataset (Fig.~\ref{Fig:Time-Independent}) as a function of the number of shots for the feature vectors $\vec{\mu}_1, \vec{\mu}_{\leq 2}$ and $ \vec{\mu}_{\leq 3}$. We see that our quantum reservoir has non-trivial third-order correlations and that the reservoir leverages these correlations to boost classification accuracy. The covariance matrix averaged over the entire spiral dataset is plotted in Fig.~\ref{fig:si:spiral_correlations2}, and the third-order correlations are plotted in Fig.~\ref{fig:si:spiral_correlations3} -- plotted as a set of 2D matrices. In the third-order correlations in particular, we can begin to pick out by eye the differences in the two classes. 

This construction generally allows us to construct feature vectors that are smaller than the probability distribution over all possible measurement trajectories, which is $2^{M}$ dimensional. However, as can be seen in Fig.~\ref{fig:si:spiral_correlations2}, there is yet redundant information even after taking only the symmetric part - specifically, that the information tends to be very local and that measurements far apart tend not to be correlated. This has the physical interpretation that while measurements are indeed correlated, even possessing higher-order correlations, this correlation tends to be local due to the finite memory of the system. This motivates us to further restrict our feature vector to only capture the essential local correlations. 

\begin{figure*}[h]
    \centering
    \includegraphics[width=0.5\textwidth]{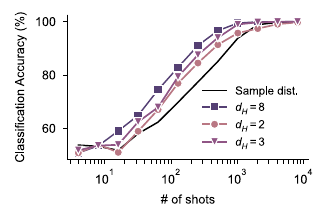}
    \caption{
    \textbf{ Spiral classification accuracy as a function of locality of up-to third-order correlations }  Classification accuracy for the spiral task using up to third-order central moments truncating correlations to only include correlations up to $d_H$. We find that we can achieve high accuracy using local third-order correlations, saturating the accuracy when only keeping correlators up to $d_H \leq 3$. The performance of using the sample distribution is also compared which performs worse than using the central moments as output features, despite containing more information (see text).   }
    \label{fig:si:spiral_accuracy_hamming}
\end{figure*}

Figure~\ref{fig:si:spiral_accuracy_hamming} compares the classification performance of feature vectors generated with up to third-order moments, where we truncate the locality of the correlations. That is, the elements of the third order central moment $(\mu_3)_{ijk}$ are set to zero if $\vert i - j \vert > d_H$ or $\vert i - k\vert > d_H$, for some integer $d_H$. We note that including third-order correlations between measurements that are up to three `sites' away nearly reproduces the classification accuracy of when you include all third-order central moments. Additionally, we compared the construction of feature vectors using truncated moments up to third-order with that of using the full sampled distribution as the feature vector and found that the former performed much better (Fig.~\ref{fig:si:spiral_accuracy_correlations}). These last two statements were found to be true for all tasks presented in this paper. For $M = 8$ measurements, the truncation of long range correlations further reduces the output feature size from 164 down to 94.

%%%%%%%%%%%%%%%%%%%%%%%%%%%%%%%%%%%%%%%%%%%%%%%%%%%%%%%%%%%%%%%%%%%%%%%%%%%%%%%%%%%%
%%%%%%%%%%%%%%%%%%%%%%%%%%%%%%%%%%%%%%%%%%%%%%%%%%%%%%%%%%%%%%%%%%%%%%%%%%%%%%%%%%%%
%%%%%%%%%%%%%%%%%%%%%%%%%%%%%%%%%%%%%%%%%%%%%%%%%%%%%%%%%%%%%%%%%%%%%%%%%%%%%%%%%%%%
\subsection*{Training the linear layer}\label{sec:si:ML:linear_layer}
\addcontentsline{toc}{subsection}{\nameref{sec:si:ML:linear_layer}}

The only component of the reservoir that was trained to fit the dataset processed by the reservoir was the linear layer applied to the feature that the physical reservoir produced. The linear layer was an $R\times C$ matrix $W_{\text{train}}$ and $C$-dimensional vector $v_{\text{train}}$ applied to the $R$-dimensional reservoir feature $x$ to get 
\begin{equation}
y=x^T W_{\text{train}}  + v_{\text{train}}
\label{e:linear layer application}
\end{equation}
where the largest of the $C$ elements of $y$ corresponded to the predicted class of the data point ($C$ is the number of classes). To train the linear layer, we chose between two different approaches: the pseudo-inverse method and back-propagation through a softmax function on the output. The two approaches optimize the linear layer over different loss landscapes. This is because our classification method is fundamentally discrete - i.e. we identify the class simply based on whichever output vector entry is the largest - so there is not a perfect correspondence between our loss  and the classification inaccuracy.

First, we will describe the pseudo-inverse method. Let $X$ be a $N\times (R+1)$ matrix consisting of $R$-dimensional reservoir features generated for $N$ training points, with a column of 1's appended (this is to compute both $W_{\text{train}}$ and $v_{\text{train}}$ at once). Let $Y$ be an $N\times C$ matrix consisting of $C-$dimensional row vectors that serve as labels for the training points such that $Y_{i,j}=1$ if $j$ corresponds to the class of the $i^{\text{th}}$ training point and zero otherwise. For an $\epsilon > 0$, we construct $W^{'}_{\text{train}}$ ($W_{\text{train}}$ appended with $v_{\text{train}}$) as : 
\begin{equation}
W^{'}_{\text{train}} = \left(X^TX+\epsilon I \right)^{-1}X^T Y
\label{e:pseudoinverse equation}
\end{equation}

In our case, the value of $\epsilon$ was swept to maximize the accuracy of the classification. In the limit of $\epsilon\rightarrow 0$, the pseudo-inverse matrix of Eq.~\ref{e:pseudoinverse equation} is provably optimal for minimizing  $|| XW_{\text{train}} - Y||_2^2$ \cite{leastsquarestextbook} up to numerical stability, and so has been a popular choice for training the linear layer at the output of reservoirs~\cite{nakajima2018reservoir,LESNREVIEW2022,Gauthier_2021, scardapane2017}. However, our goal was to classify the input signals based on the \emph{largest} element of the final output vector.  Consequently, the linear layer that resulted in the lowest mean squared error with our labels was not always the linear layer that gave us the best accuracy. 

For this reason, we also used a second training method for our linear layer. This approach used softmax, a popular choice for classifiers in neural networks \cite{banerjee2020} and back-propagation using the automatic differentiation package from PyTorch~\cite{pytorch2019}. Training through back-propagation with an optimizer is now necessary since an exact analytic solution to minimize the loss no longer exists, unlike the case of the pseudo-inverse. In this approach, the prediction vector $y$ from Eq.~\ref{e:linear layer application} is passed through the ``Softmax" activation function:
\begin{equation}
(y_{\text{prediction}})_i = \frac{\exp(y_i)}{\sum_{j=1}^C \exp(y_j)} 
\label{e:softmax}
\end{equation}
We then computed the mean squared error between the resulting $y_{\text{prediction}}$ and the label for the training point that produced the underlying reservoir feature $x$. Finally, we used back-propagation to compute the gradient for our linear layer. The linear layer was then updated using the ADAM optimizer~\cite{ADAMpaper} with the default settings of $\beta_1=0.9,\beta_2=0.999$ and a learning rate of $0.01$. For our reservoirs, we tried both methods of training the linear layer and used whichever yielded the best accuracies. Empirically, we found that while pseudo-inverse training was better in some cases, training the linear layer with back-propagation often yielded quite large accuracy advantages over pseudo-inverse.

%%%%%%%%%%%%%%%%%%%%%%%%%%%%%%%%%%%%%%%%%%%%%%%%%%%%%%%%%%%%%%%%%%%%%%%%%%%%%%%%%%%%
%%%%%%%%%%%%%%%%%%%%%%%%%%%%%%%%%%%%%%%%%%%%%%%%%%%%%%%%%%%%%%%%%%%%%%%%%%%%%%%%%%%%
%%%%%%%%%%%%%%%%%%%%%%%%%%%%%%%%%%%%%%%%%%%%%%%%%%%%%%%%%%%%%%%%%%%%%%%%%%%%%%%%%%%%
\section{Supplementary information machine learning tasks}\label{sec:si:tasks_si}

\subsection*{Classification of Radio-Frequency signals}\label{sec:si:tasks_si:rfml}
\addcontentsline{toc}{subsection}{\nameref{sec:si:tasks_si:rfml}}

In this section, we discuss about the algorithm for generating the dataset for the classification of digital modulation schemes on radio signals. The digital modulation scheme involves encoding sequences of binary values into the amplitude and phase of a radio signal for a fixed duration. The number of binary values encoded depends on the modulation scheme. For example, for BPSK (binary phase shift key), each symbol (change in property of the signal) encodes one bit of information. For $32$QAM (quadrature amplitude key), there are $32$ possible values, which allows each symbol to contain $5$ bits of information. For this task, we keep the symbol rate fixed across all the tasks. Moreover, the pulses generated by the arbitrary waveform generator (AWG) all occur at the baseband frequency. This signal is then upconverted to the frequency of the cavity before sent to the device. To generate the set of possible sequences, we randomly select each symbol with equal probability. This corresponds to the case of each possible binary string of digits encoded to be equally likely. Due to memory constraints on the AWG, we cannot output a continuous encoded signal for long durations, corresponding to the regime of large samples of the reservoir. We circumvent this constraint by realizing that, for this task, there are no correlations in the encoded binary digit sequence (since each symbol is equally likely). Therefore, the probability of a long binary digit sequence can be correctly emulated by sampling multiple short binary digit sequences and concatenating them together. For this task, we can simply achieve this by generating a signal with eight symbols, which is the number of symbols enter our QRC before its state is completely reset.

%%%%%%%%%%%%%%%%%%%%%%%%%%%%%%%%%%%%%%%%%%%%%%%%%%%%%%%%%%%%%%%%%%%%%%%%%%%%%%%%%%%%
%%%%%%%%%%%%%%%%%%%%%%%%%%%%%%%%%%%%%%%%%%%%%%%%%%%%%%%%%%%%%%%%%%%%%%%%%%%%%%%%%%%%
%%%%%%%%%%%%%%%%%%%%%%%%%%%%%%%%%%%%%%%%%%%%%%%%%%%%%%%%%%%%%%%%%%%%%%%%%%%%%%%%%%%%
\subsection*{Classification of noisy signals}\label{sec:si:tasks_si:noise_ml}
\addcontentsline{toc}{subsection}{\nameref{sec:si:tasks_si:noise_ml}}

\begin{figure*}[h]
    \centering
    \includegraphics[width=0.6\textwidth]{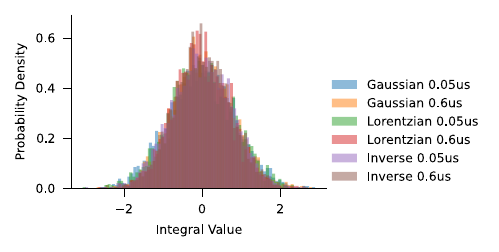}
    \caption{\textbf{Histogram of integrated value of the input signal for the noise classification task.} In this work, we enforced a normalization condition on the amplitude of the filter functions. We set the normalization such that the long time integral of the signal corresponds to a value with zero mean and the same standard deviation. This is visually seen from the probability density function from the dataset of the classes of noisy signals. We do this to ensure that any reservoir which simply integrated the signal, before applying a non linear kernel, cannot classify the different signals. The fact that our reservoir is able to solve this task can therefore be associated with the continuous-time processing by the cross-Kerr interaction between the qubit and the cavity. Enforcing this normalization is mathematically equivalent to setting the DC component of the filter functions to be the same value in frequency space.}
    \label{fig:si:noise_integration}
\end{figure*}

To generate the dataset describing the task of classifying noisy signals using the QRC (see Fig.~\ref{Fig:Noise_ML}), we start with emulating white noise. At each time step of the sampling rate of the AWG, we choose a value for in-phase and quadrature signals uniformly between the unit interval (up to an overall normalization). While this is limited to the sampling rate of the AWG (around $2\times 10^9$ samples per second), this is much larger than any relevant time scale of the experiment. Therefore the approximation of broadband white noise is appropriate to describe the effect of the signal on the system. We then apply ``kernels" as a convolution in the time domain to each need seed of the white noise generated signal. This can also be thought of a bandpass filtering function in frequency domain. The classification task is then to identify the kernel. Each kernel is defined by a time domain function. The only hyper-parameter to describe each kernel is the overall scaling value. In this work, we set the DC component of this kernel in frequency domain to be the same for all classes (set to unit value without loss of generality). In the time domain, this corresponds to scaling the amplitude such that the area enclosed by the filter function in time is the same for all functions.. We do this to make sure that a direct integration of the signal over a time domain much longer than the correlation length introduced by the kernel, cannot distinguish the signals from each other (see Fig.~\ref{fig:si:noise_integration}). The above normalization ensures the random variable associated with this integrated value is the same for all distributions. Therefore, this ensures that any ability of the reservoir to classify the signals arises intrinsically from its computational capacity to distinguish short-time correlations (in this work we choose a correlation time scale of $50$ns and $600$ns, with kernel functions of Gaussian, Lorentzian, and the Inverse function: generating a total of $6$ classes.).

%%%%%%%%%%%%%%%%%%%%%%%%%%%%%%%%%%%%%%%%%%%%%%%%%%%%%%%%%%%%%%%%%%%%%%%%%%%%%%%%%%%%
%%%%%%%%%%%%%%%%%%%%%%%%%%%%%%%%%%%%%%%%%%%%%%%%%%%%%%%%%%%%%%%%%%%%%%%%%%%%%%%%%%%%
%%%%%%%%%%%%%%%%%%%%%%%%%%%%%%%%%%%%%%%%%%%%%%%%%%%%%%%%%%%%%%%%%%%%%%%%%%%%%%%%%%%%
\section{Simulation of the quantum reservoir}\label{sec:si:simulation}

%%%%%%%%%%%%%%%%%%%%%%%%%%%%%%%%%%%%%%%%%%%%%%%%%%%%%%%%%%%%%%%%%%%%%%%%%%%%%%%%%%%%
%%%%%%%%%%%%%%%%%%%%%%%%%%%%%%%%%%%%%%%%%%%%%%%%%%%%%%%%%%%%%%%%%%%%%%%%%%%%%%%%%%%%
%%%%%%%%%%%%%%%%%%%%%%%%%%%%%%%%%%%%%%%%%%%%%%%%%%%%%%%%%%%%%%%%%%%%%%%%%%%%%%%%%%%%

\subsection*{Introduction}\label{sec:si:simulation:introduction}
\addcontentsline{toc}{subsection}{\nameref{sec:si:simulation:introduction}}

\begin{figure*}[h]
    \centering
    \includegraphics[width=\textwidth]{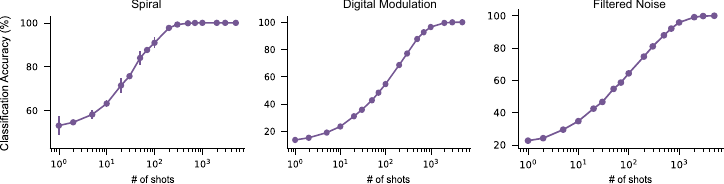}
    \caption{\textbf{Classification accuracies obtained from numerical simulations of the QRC.} Numerical simulations of the quantum reservoir can help guide the expected performance of the system in experiment. Here, we simulate the three main tasks considered in the paper: Spiral (a time-independent input with the goal of classifying the arms of the spiral), Digital Modulation (a slow-varying, time-dependent signal with the goal of identifying the digital modulation scheme used in the signal), Correlated Noise (a fast-varying, time-dependent signal with the goal of classifying the ``kernel" of the correlation function). }
    \label{fig:si:simulation:cumulants}
\end{figure*}

Classical simulations of the QRC can provide insight into the expected computational capacity in experiment. For our work, classical simulations of the dynamics of the reservoir were primarily performed with the aid of QuTiP~\cite{Johansson_2012}. The algorithm to estimate the classification accuracy for a given task then follows the same technique used in experiment, with training and testing datasets on the measurement outcomes of the simulation. We implement the Hamiltonian in Eq~\ref{eq:Hamiltonian}, by approximating the transmon as a qubit, and introducing a finite dimensional Fock truncation to the cavity subspace. It is important to ensure that the Fock truncation does not introduce any spurious effects, for it can be a source of non-physical non-linearities in the system. For example, a linear cavity, treated as a harmonic oscillator, only performs a linear transformation on an incoming analog radio frequency signal. However, if in simulation, the support of the state of the cavity exceeds the Fock truncation of the simulation, numerical errors introduce non-Gaussian states in the cavity mode. Such effects will depend non-linearly on the input, and hence can effectively act as a ``good" (but of course unphysical) reservoir! To ensure this doesn't happen in simulation, at every step of the unitary evolution, we monitor the probability of the wavefunction on the largest Fock state in the simulation. If this value goes above $1\%$ during the simulation, a warning is raised, and the results of the simulations are discarded. 

To make the simulations efficient, we make certain assumptions on the quantum system. Firstly, we treat the reservoir controls of qubit rotations and conditional displacements with a ``gate"-based unitary. However, to take into account the analog, continuous dynamical evolution implemented by the cross-Kerr interaction term in the Hamiltonian, the interval of the input into the system is implemented with the full time-dependent Hamiltonian evolution (using QuTiP's ``mesolve" functions). Finally, another approximation we make (in favor of simulation speed) is ignoring decoherence effects. To ensure this approximation is valid, we performed simulation with the stochastic wavefunction approach with photon loss and qubit dephasing rates measured in experiments~\cite{PhysRevLett.68.580}. We obtain differences in expected classification accuracies within error bars (which are obtained from different datasets from repeated simulations). This also gives us confidence that the role of decoherence in the system plays a minimal role in the computational capacity of the reservoir. The results of the simulations with the third central moment are plotted in Fig.~\ref{fig:si:simulation:cumulants}. Interestingly, the performance as a function of the number of samples agrees to experiment within the same order of magnitude. This gives a good estimate for the experimental time required to produce a classification accuracy versus shots curves in experiments. For all three tasks, the reservoir approaches $100\%$ accuracy with sufficient samples or integration time of the input. 

%%%%%%%%%%%%%%%%%%%%%%%%%%%%%%%%%%%%%%%%%%%%%%%%%%%%%%%%%%%%%%%%%%%%%%%%%%%%%%%%%%%%
%%%%%%%%%%%%%%%%%%%%%%%%%%%%%%%%%%%%%%%%%%%%%%%%%%%%%%%%%%%%%%%%%%%%%%%%%%%%%%%%%%%%
%%%%%%%%%%%%%%%%%%%%%%%%%%%%%%%%%%%%%%%%%%%%%%%%%%%%%%%%%%%%%%%%%%%%%%%%%%%%%%%%%%%%

\subsection*{The advantage of continuous-time continuous-variable QRCs over discrete-time qubit-based QRCs}\label{sec:si:simulation:continuous}
\addcontentsline{toc}{subsection}{\nameref{sec:si:simulation:continuous}}

\begin{figure*}[h]
    \centering
    \includegraphics[width=0.9\textwidth]{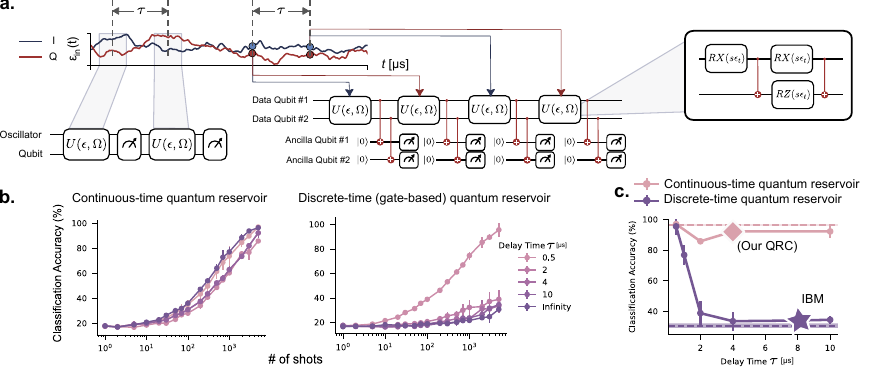}
    \caption{\textbf{Simulations comparing the performance of a continuous-time quantum reservoir (our work), and a discrete-time qubit-based quantum reservoir, based on a recently implemented protocol~\cite{yasuda2023quantum}.} \textbf{(a.)} Schematic reservoir protocol for our experiment (on the left), versus that introduced in ~\cite{yasuda2023quantum}, involving two qubits as the data qubits, and two as the ancillas. The reservoir consists of single qubit rotations interleaved with CNOT gates between the data qubits and data and ancilla qubits. In this simulation, we simulate the performance of the two reservoirs as a function of the delay $\tau$ between two durations of the input. Such a delay can be introduced by the finite pulse durations and latencies introduced by the FPGA. \textbf{(b.)} Classification accuracy curves for the two reservoirs as a function of shots, for different values of delays. The continuous time analog reservoir is much more robust to delays between inputs compared to the discrete time reservoir implementation. \textbf{(c.)} Plot of accuracies at $5000$ shots for the two reservoir implementations as a function of delay time. Experimentally relevant times include $4\mu s$ for our experiment, and around $8 \mu s$~\cite{hua2023exploiting} for the experimental realization of~\cite{yasuda2023quantum} on an IBM quantum computer.}
    \label{fig:si:simulation:digital_comparison}
\end{figure*}

In this section, we benchmark the performance of our continuous-time-continuous variable QRC in comparison with other hardware implementations of reservoirs. To highlight the benefit of our QRC in processing time varying input signals, we compare the simulation of our reservoir with that of a recent QRC scheme involving repeated measurements on a multi-qubit based superconducting circuit quantum system~\cite{yasuda2023quantum}. For this comparison, we simulate the expected performance of both systems on the task of classifying different noise signals with the classes described in Fig.~\ref{Fig:Noise_ML}. While our reservoir can naturally interface with analog signal, this is not the case with the protocol introduced in~\cite{yasuda2023quantum}. For this simulation, the signal is sampled at discrete times and input to the system as a scalar parameter (one for the in-phase value and one for quadrature value). To highlight the advantage of our QRC, we slightly modify the task introduced in the Fig.~\ref{Fig:Noise_ML}. Here, we normalize the six filter functions such that the integral of the filter function in frequency domain is kept constant. We do this such that the standard deviation associated with the distribution of the sampled signal is the same across all signals. The only information distinguishing the signals is in the correlation between two close samples in time. To elucidate this reasoning, we simulate the performance of the two reservoirs as a function of the time duration in between two samples of the signal (in the case of the discrete qubit based reservoir) and integration windows (for our analog reservoir). Such a finite duration can arise from finite-pulse durations of reservoir protocols, qubit-measurement times, and the finite latency introduced by the classical FPGA processor. For example, for our experiment, this time is around $4\mu s$, mostly arising from the measurement of the qubit and the parity of the cavity. In experiment (Fig.~\ref{Fig:Noise_ML}), we had generated and timed the input wave forms such that the delay between inputs is essentially $0 \mu s$. For a typical IBM quantum device with mid-circuit measurement, the protocol used in Ref.~\cite{yasuda2023quantum}, the finite latency can be estimated to be around $8\mu s$~\cite{hua2023exploiting}. The protocol for the discrete-time quantum reservoir is designed to only act on real-valued input signals. However, for a continuous signal in the rotating frame, we have both the in-phase and quadrature values. For the experimental quadrature, these values correspond to displacements on the oscillator in orthogonal directions. To extend the scheme presented in~\cite{yasuda2023quantum}, we do the following minimal change: we interleave the between sample points of in-phase and quadrature values. We could have chosen these points with the delay of $\tau$ in between each. However, this might have had the effect of introducing twice the delay compared to the continuous time reservoir. Therefore, we chose the relaxed constraint of the input such that both the in-phase and quadrature values are chosen at the same point, with just a delay in between two different in-phase and quadrature points.

%%%%%%%%%%%%%%%%%%%%%%%%%%%%%%%%%%%%%%%%%%%%%%%%%%%%%%%%%%%%%%%%%%%%%%%%%%%%%%%%%%%%
%%%%%%%%%%%%%%%%%%%%%%%%%%%%%%%%%%%%%%%%%%%%%%%%%%%%%%%%%%%%%%%%%%%%%%%%%%%%%%%%%%%%
%%%%%%%%%%%%%%%%%%%%%%%%%%%%%%%%%%%%%%%%%%%%%%%%%%%%%%%%%%%%%%%%%%%%%%%%%%%%%%%%%%%%

\subsection*{Comparison to other reservoirs}\label{sec:si:simulation:reservoirs}
\addcontentsline{toc}{subsection}{\nameref{sec:si:simulation:reservoirs}}

\begin{figure*}[h]
    \centering
    \includegraphics[width=\textwidth]{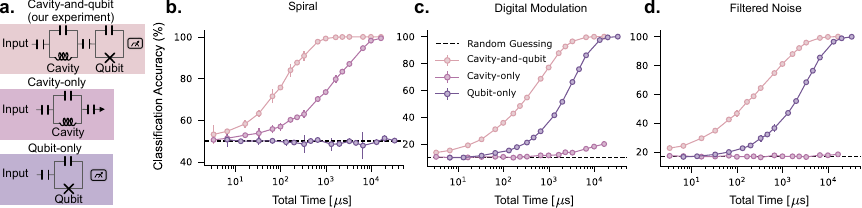}
    \caption{\textbf{Simulations of the performance of the cavity-and-qubit reservoir system with its components: just a cavity reservoir and just a qubit reservoir.} \textbf{(a.)} The circuit diagrams of the three quantum reservoir simulated here. 1) The circuit diagram of the experimental QRC in this work: a cavity coupled to a qubit. The input is interfaced to the cavity, and the state of the qubit is measured using the standard dispersive readout technique with a readout resonator. 2) The circuit diagram of a cavity as a reservoir. In this case, the natural output of the cavity is the transmitted signal. 3) The circuit diagram of a qubit reservoir. In this case the input is coupled instead to the qubit. \textbf{(b.)} Simulation of the classification accuracy of the three different reservoirs for the three tasks considered in the paper. As a fair comparison, the output feature vector dimension is kept constant for all three reservoir. The experimental setup drastically outperforms either of its components, just a cavity (in \textbf{(c.)} and just a qubit (in \textbf{(d.)}), which highlights the important role of entanglement in classification accuracy.}
    \label{fig:si:simulation:reservoir}
\end{figure*}

A cavity coupled to a qubit is a hardware efficient quantum system to perform reservoir computing on analog signals. In this section, we motivate this by simulating the performance of other natural choices of quantum reservoirs: a single qubit, and a single cavity. The protocols for these systems are inspired by what one can naturally perform in experiment. To make a reservoir with a cavity, we couple the input into the cavity (as is the case for the experimental design). To readout the cavity, we perform a transmission style Homodyne measurement, which infers the mean field value of the cavity. This is a continuous form of measurement, where the output feature is a time dependent radio frequency signal at the frequency of the cavity mode. Since the cavity is always in a coherent state, the output time trace is linearly dependent on the incoming signal. For a fair comparison, we only use a handful of values from the time trace (as many as the number of measurements in the experiment). While this might seem restrictive, we process this via the same method as the case of the experimental reservoir, by computing the functional definition of the central moments. This does not necessarily make sense for this protocol, since the outputs do not correspond to samples from a discrete probability distribution, but can nevertheless introduce non-linearities in the representation of the feature vector. These non-linearities can improve the performance of the reservoir beyond a linear layer. This is observed for the case of time-independent Spiral classification, where the cavity reservoir performs better than random. This performance is solely due to the ``post-processing" of the output of the reservoir we adopt for our experiment. However, for time dependent tasks, the performance is hardly better than random.

Another natural candidate is a single qubit reservoir. For this case, we directly interface the signal to the qubit. A qubit is able to naturally represent non-linear functions of the input, which can be intuitively seen by visualizing the action of a qubit rotation on a Bloch sphere. As a fair comparison, we choose the same qubit reservoir controls in experiment, which involve qubit pulses before, during and after the continuous input. The output is a string of binary outcomes of qubit measurements, which can be done experimentally with the use of a readout resonator. Each reservoir of the qubit lasts twice as long as the experimental QRC to obtain the same feature vector size. This is then processed the same way as the cavity-qubit coupled reservoir, before applying a trained linear layer. Interestingly, the qubit fails to perform better than random for the Spiral task. On the other hand, it is able to reach near $100\%$ accuracy for the time dependent signal classification tasks. The ability for even a single qubit to successfully perform a many-class classification task is illuminating at the remarkable processing capabilities of reservoir. However the total input signal required can be more than order of magnitude longer compared to that for the experimental QRC to achieve the same accuracy. The ability of a cavity coupled to a qubit system to perform significantly better than either of its components provides a clear picture of the important role entanglement can play.

%%%%%%%%%%%%%%%%%%%%%%%%%%%%%%%%%%%%%%%%%%%%%%%%%%%%%%%%%%%%%%%%%%%%%%%%%%%%%%%%%%%%
%%%%%%%%%%%%%%%%%%%%%%%%%%%%%%%%%%%%%%%%%%%%%%%%%%%%%%%%%%%%%%%%%%%%%%%%%%%%%%%%%%%%
%%%%%%%%%%%%%%%%%%%%%%%%%%%%%%%%%%%%%%%%%%%%%%%%%%%%%%%%%%%%%%%%%%%%%%%%%%%%%%%%%%%%

\subsection*{Multi-qubit reservoirs}\label{sec:si:simulation:multi}
\addcontentsline{toc}{subsection}{\nameref{sec:si:simulation:multi}}

\begin{figure*}[h]
    \centering
    \includegraphics[width=0.75\textwidth]{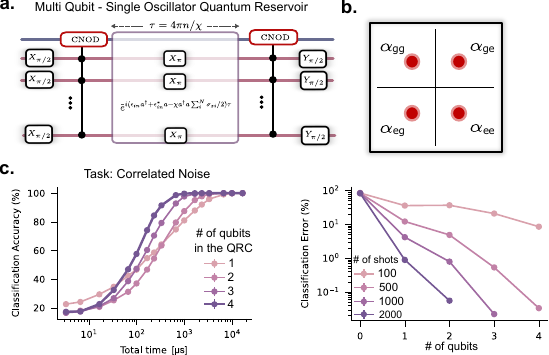}
    \caption{\textbf{Exploring the computational capacity of more complex reservoirs, involving a single cavity mode coupled to multiple qubits.} \textbf{(a.)} The reservoir protocol, before measurement. The protocol is inspired by naturally extending the protocol of the experiment. The state of each qubit, along with the parity of the cavity, is sampled afterwards. \textbf{(b.)} Schematic illustration of the state of the cavity after the generalized qubits conditioned cavity displacement implemented in this reservoir, for the case of two qubits. Here the cavity is displaced by a unique value of each $2^N$ combinations of qubit possibilities, for a reservoir with N qubits. \textbf{(c.)} To illustrate the computational capacity of such reservoir, we simulate the quantum system to obtain the classification accuracy as a function of the duration of signal received for the task of classifying correlated noise signals. Increasing the number of qubits generally increases the performance of the reservoir. For different values of shots of the reservoir, which corresponds a total time of input, we plot the classification error as a function of qubits. The case of zero qubits in the reservoir corresponds to just a cavity reservoir.}
    \label{fig:si:simulation:multi}
\end{figure*}

Quantum reservoir computing is a promising paradigm in the NISQ era. It is therefore interesting to consider the potential benefits in performance with larger devices, which are within reach of today's experimental capabilities. As a natural extension of our quantum reservoir, we consider a scenario of one continuous variable cavity mode dispersively coupled to multiple qubits. For simplicity, we assume the dispersive strength of each qubit to the cavity is the same. To motivate the capacity of such a reservoir, we simulate the system for up to four qubits to estimate the classification accuracy for the task of identifying correlated noise signals. The unitary protocol is illustrated in Fig~\ref{fig:si:simulation:multi} (a.). The protocol begins with a $\pi/2$ pulse on each qubit, which brings the state of the qubit onto the equator of the Bloch sphere. To entangle the qubits with cavity, we simulate the action of a generalized, multi-qubit-conditioned cavity displacement. This involves a displacement on the cavity, whose value is different for each of $2^N$ N-qubit possibilities. An example of the action of this operator is depicted in Fig~\ref{fig:si:simulation:multi} (b.) for the case of two qubits. Here, there are four possible qubit states, and each is associated with a displacement value of the four corners of the square. The displacement values were chosen somewhat arbitrarily, but serve to illustrate an efficient multi-component entanglement. For the case of two qubits, the real and imaginary components of the displacement where either $\pm 0.5$. For the case of three qubits, there are eight total possible states. The set of displacements chosen form a three-by-three grid state, ranging between $\pm 1$, excluding the center of this grid (which is centered at the origin). For the case of four qubits, a four-by-four grid uniformly distributed between $\pm 1.5$ covers all sixteen possibilities. The correspondence between qubit states and displacements was somewhat arbitrary - the motivation is that even without much design choice, a reservoir can successfully implement machine learning! For these simulations, each position of the grid is associated with a decimal value, increasing sequentially from left to right, starting from the top left and progressing towards the bottom right (starting with zero). This decimal value is the decimal representation of the qubit-state bit string that the displacement is conditioned on. 

After the multi-qubit entangling conditional displacement gate, the cavity is subject to the input. The dynamics of the system are influenced by the cavity-qubit coupled dispersive interactions, where the interaction strength is the same between the cavity and all qubits and set to that of the experiment. Like the experimental QRC, each qubit is flipped with a $\pi$ pulse in the middle of the input. The protocol ends with the same conditional displacement, before a $\pi/2$ pulse. The output of the reservoir is the measurement of each qubit, along with the parity measurement of the cavity. This protocol is repeated four times, to match the experimental protocol as much as possible. 

Fig~\ref{fig:si:simulation:multi} (c.) is the classification accuracy as a function of the total time of input signal received for the reservoirs with different number of qubits, for the task of noise classification. While they all achieve essentially $100\%$ accuracy, the total time required to achieve this accuracy drops significantly with increasing number of qubits. Other than the single qubit reservoir (the experimental protocol), the performance of the reservoir is similar both at the low and high signal duration regime, differing only in the intermediate regime. The reason for the difference in behavior of the performance for the case of cavity coupled to a single qubit is the slight change in reservoir protocol. To accurately account for the experimental protocol, the state of the qubit is determined by the outcome of the parity measurement of the cavity. This was not implemented in the simulations for multiple qubits. This ends up improving the performance of this reservoir for this task in the low signal duration regime. However, in the higher signal duration regime, increasing the number of qubits increases the accuracy. 

The classification error as a function of number of qubits in the reservoir is plotted in Fig~\ref{fig:si:simulation:multi} (d.), including the case for just a cavity (zero number of qubits in the reservoir), for a select number of total shots of the entire reservoir. Very crudely, the error in classification seems to reduce exponentially with every additional qubit in the reservoir.

%%%%%%%%%%%%%%%%%%%%%%%%%%%%%%%%%%%%%%%%%%%%%%%%%%%%%%%%%%%%%%%%%%%%%%%%%%%%%%%%%%%%
%%%%%%%%%%%%%%%%%%%%%%%%%%%%%%%%%%%%%%%%%%%%%%%%%%%%%%%%%%%%%%%%%%%%%%%%%%%%%%%%%%%%
%%%%%%%%%%%%%%%%%%%%%%%%%%%%%%%%%%%%%%%%%%%%%%%%%%%%%%%%%%%%%%%%%%%%%%%%%%%%%%%%%%%%

\section{Theoretical analysis of the expressivity of our QRC for time-independent signals}\label{sec:si:expressivity}

The ability of the QRC to perform better than an optimal linear layer on the input lies in the reservoir's ability to express many non-linear functions of the input---its expressivity. Here, we quantitatively characterize the class of functions which can be represented by the oscillator component of the QRC for a time-independent input. In this regime, the input can be represented by two variables: the values of the in-phase and quadrature components. The output feature vector from the QRC is then a function of these two variables.

\noindent We denote $\alpha = |\alpha| e^{i \phi_{\alpha}}$,
$\beta = \vert \beta \vert e^{i \phi_{\beta}}$, and set $\vert \alpha \vert = 1/2$. Choosing different values of $\phi_{\alpha}$ gives rise to different output features of the QRC.
In this experiment, we pick $\phi_{\alpha} \in \{0, \pi/2\}$, but in principle, one can add to the feature vector with more choices of $\phi_{\alpha}$. For example, one can choose $\phi_{\alpha} \in \{0, \omega, 2 \omega, \dots, (r-1) \omega\}$ where $\omega = \frac{2\pi}{r}$. The final output after the linear layer is an arbitrary linear combination of all the $p_{\alpha}(\beta)$ functions.

Intuitively, the larger $r$, the more expressive the function space spanned by these features. Furthermore, the higher-order central moments allow the output feature vector to represent powers of this probability: $p_{\alpha}(\beta)^n$, for moments up-to the nth-order. We have shown that the qubit measurements extract the phase information of the input complex number $\beta$.
Below we will focus on the oscillator parity measurement which is sensitive to the magnitude of $\beta$. Recall that the post-measurement (unnormalized) state of the cavity can be described by a sequence of alternating displacements and parity measurements (Eq.~\ref{eq:si:repeated_cavity}):

\begin{equation}
    \vert \Psi_{\vec{x}}(\beta) \rangle = P_{x_M} D(\beta) \cdots P_{x_2} D(\beta) P_{x_1} D(\beta) \vert 0 \rangle,
    \label{eq:si:ket-cav}
\end{equation}

\noindent where $P_{x_i}$ is the projector of the $i$-th parity measurement with outcome $x_i \in \{0,1\}$, with `0' standing for `even' and `1' for `odd'. That is, $P_{x_i} = \frac{I + (-1)^{x_i} \Pi}{2}$, where $\Pi = (-1)^{a^{\dagger} a}$.
The corresponding probability of obtaining $\vec{x} = (x_1, x_2, \dots, x_M)$ as the sequence of measurement results given the input $\beta$ is
\begin{equation}
    \Pr \left[ \vec{x} | \beta \right] = \langle \Psi_{\vec{x}}(\beta) | \Psi_{\vec{x}}(\beta) \rangle.
\end{equation}

To obtain a simplified expression for $\Pr \left[ \vec{x} | \beta \right]$, we will make use of the following formula:
\begin{equation}
    P_x D(\beta) P_y = \frac{D(\beta) + (-1)^{x \oplus y} D(-\beta)}{2} P_y, \quad \forall x \in \{0,1\}, \forall y \in \{0,1\},
    \label{eq:si:PxDPy}
\end{equation}
which is an easy application of the commutation relation $\Pi D(\beta) = D(-\beta) \Pi$, with the latter being derived from $\Pi a = -a \Pi$. Using Eq.~\ref{eq:si:PxDPy}, we can remove all the explicit parity projectors in Eq.~\ref{eq:si:ket-cav}:
\begin{equation}
    \ket{\Psi_{\vec{x}}(\beta)} = \left( \prod_{i=1}^{M} \frac{D(\beta) + (-1)^{x_i \oplus x_{i-1}} D(-\beta)}{2} \right) \ket{0},
\end{equation}
where for notational simplicity we have prepended the bit-string $\vec{x}$ by $x_0 \equiv 0$. Note that the order of the product does not matter since the terms commute with each other. It follows that:

\begin{equation}
    \begin{aligned}
    \Pr \left[ \vec{x} \vert \beta \right] & = \langle 0 \vert \left( \prod_{i=1}^{M} \frac{D(-\beta) + (-1)^{x_i \oplus x_{i-1}} D(\beta)}{2} \right) \left( \prod_{i=1}^{M} \frac{D(\beta) + (-1)^{x_i \oplus x_{i-1}} D(-\beta)}{2} \right) \vert 0 \rangle \\
    &= \langle 0 \vert \left( \prod_{i=1}^{M} \left[ \frac{1}{2} + (-1)^{x_i \oplus x_{i-1}} \frac{D(2\beta) + D(-2\beta)}{4} \right] \right) \vert 0 \rangle.
    \end{aligned}
    \label{eq:si:Pr-vec-x}
\end{equation}

There are multiple methods to encode the measurements of the QRC. Representing every binary string of measurement outcomes as the feature, the output of the QRC are all the probabilities $\left\{ \Pr \left[ \vec{x} | \beta \right] \right\}_{\vec{x} \in \{0,1\}^M}$.
From Eq.~\ref{eq:si:Pr-vec-x}, it is not hard to see that when regarded as functions of $\beta$, these $2^M$ features linearly span a $(M+1)$-dimensional function space that has the following basis functions:
\begin{equation}
    f_k(\beta) := \bra{0} D(2k\beta) \ket{0} = e^{-2k^2 \vert \beta \vert^2}, \quad k=0,1,2,\dots,M.
\end{equation}
Therefore, the set of all functions realizable by the QRC combined with the linear layer is
\begin{equation}
    \left\{ c_0 f_0(\beta) + c_1 f_1(\beta) + \cdots + c_M f_M(\beta) : c_0, c_1, \dots, c_M \in \mathbb{R} \right\}.
\end{equation}

Given the large redundancy of the output feature encoding manifested above, a compact representation can be the centralized moments $\mu_{i_1, i_2, \dots, i_k}(\beta) := \mathop{\mathbb{E}} \left[ \left( x_{i_1} - \mathop{\mathbb{E}}[x_{i_1}] \right) \left( x_{i_2} - \mathop{\mathbb{E}}[x_{i_2}] \right) \cdots \left( x_{i_k} - \mathop{\mathbb{E}}[x_{i_k}] \right) \right]$. These feature functions contain terms like $\mathop{\mathbb{E}}[x_{1}] \mathop{\mathbb{E}}[x_{2}]$, $\mathop{\mathbb{E}}[x_{1}]^2$, $\mathop{\mathbb{E}}[x_{1}] \mathop{\mathbb{E}}[x_{2}] \mathop{\mathbb{E}}[x_{3}]$, and so on. In particular, for any $k$, $\mathop{\mathbb{E}}[x_{1}]^k = \left( \frac{1}{2} - \frac{e^{-2|\beta|^2}}{2} \right)^k$ can be written as a linear combination of centralized moments of order less than or equal to $k$. It follows that the QRC using at most $k$-th order centralized moments combined with the linear layer can realize (but not limited to) the following vector space of functions:

\begin{equation}
    \mathcal{H}_\mathrm{parity} := \left\{ c_0 + c_1 e^{-2 \vert \beta \vert^2} + c_2 \left( e^{-2\vert \beta \vert^2} \right)^2 + \cdots + c_k \left( e^{-2 \vert \beta \vert ^2} \right)^k: c_0, c_1, \dots, c_k \in \mathbb{R} \right\}.
\end{equation}

Note that $\mathcal{H}_\mathrm{parity}$ is exactly the set of all degree-$k$ polynomials in the variable $w \equiv e^{-2 \vert \beta \vert^2}$.
Suppose that in some classification task, the magnitude of the input has an upper bound, say, $\vert \beta \vert \le 1$, then $w$ takes value in the closed interval $[e^{-2}, 1]$. By the Stone–Weierstrass theorem, in the limit $k \to \infty$, $\mathcal{H}_k$ approximates all continuous functions of $w$ on $[e^{-2}, 1]$, and hence all continuous functions of $\vert \beta \vert$ on $[0,1]$.

\section{Leaky Echo State Networks (LESN)}
\label{sec:si:lesn}

\subsection*{Background}\label{sec:si:lesn:background}
\addcontentsline{toc}{subsection}{\nameref{sec:si:lesn:background}}
Leaky echo state networks~\cite{JAEGER2007} are a generalization of echo state networks (ESN)~\cite{jaeger2001echo} that were found to outperform their parent design in prediction and classification of slow dynamic systems, noisy time series and time-warped dynamic patterns~\cite{LESNREVIEW2022}. Given a sequence of inputs $\{u_n\}_{n=1}^N, u_n\in\mathbb{R}^D$, the state of the LESN reservoir after the $n^{\text{th}}$ input $u_n$, $x_n$, is given by the following equation:
\begin{equation}
    x_n = (1-a\gamma)x_{n-1} + \gamma f\left(W_{\text{in}} u_n + W_{\text{res}}x_{n-1}\right).
\label{e:lesn equation}
\end{equation}
Here, $a,\gamma $ are fixed hyper-parameters in $[0,1]$, and $f$ is a nonlinear activation function. $W_{\text{in}}$ is the $R\times D$ ``encoding" matrix whose elements are selected uniformly at random from the interval $[-w_{in},w_{in}]$, where $D$ is the dimension of the input, $R$ is the dimension of the reservoir, and $w_{in}$ is a fixed hyper-parameter. $W_{\text{res}}$ is the $R\times R$ ``reservoir" matrix. This matrix is constructed by first generating a matrix $W_R$, which is a random matrix whose elements are chosen to be zero with probability $1-p_s$ and a number sampled uniformly from the interval $[-1,1]$ with probability $p_s$. The largest-magnitude singular value of this matrix, $\lambda_{\text{max}}(W_R)$ is computed, and the reservoir matrix $W_{\text{res}}$ is defined as:
\begin{equation}
W_{\text{res}} = \frac{\rho}{\left| \lambda_{\text{max}}(W_R) \right|}W_R
\label{e:reservoir matrix}
\end{equation}
where $\rho$ is a fixed scaling hyper-parameter. Finally, the $n^{\text{th}}$ output of the reservoir, $y_n$, is given by
\begin{equation}
    y_n = W_{\text{train}} x_n,
\label{e:lesn output}
\end{equation}
where $W_{\text{train}}$ is a $C\times R$ trainable linear layer, where $C$ is the dimension of the desired output vector.

\subsection*{Digital reservoir comparison}\label{sec:si:lesn:comparison}
\addcontentsline{toc}{subsection}{\nameref{sec:si:lesn:comparison}}

\begin{figure}
    \centering
    \includegraphics[width=0.5\textwidth]{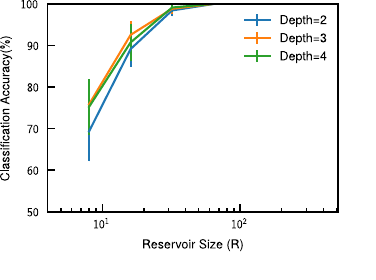}
    \caption{Mean spiral classification accuracies and their standard deviations over 100 randomly generated LESN's. 
    % Hyper-parameters for all reservoir sizes $R$ (8,16,32,64,128,256) and depths (2,3,4) are specified in Tab.~\ref{table:si:hyperparam table}.
    }
    \label{fig:si:lesn_spiral}
\end{figure}

As a way of benchmarking the computational capacity of our physical reservoir, we compared it to the performance of a \emph{digital} reservoir - an LESN - at varying widths and depths. We focused on the accuracy of classifying the spiral, since this is the most direct point of comparison, as the goal was to classify individual points of a signal rather, than multiple separate signals per-shot as with the time-dependent case. Here, for a depth of $N$, we sent in $N$ identical two-dimensional data points $(x,y)$ (so $D=2$) corresponding to the $I$ and $Q$ components of the signal that our experimental reservoir is meant to process, i.e. the spiral point coordinates. We used the rectified linear unit (ReLU) as our nonlinear activation function. Traditionally, sigmoid or tanh activation functions are used for LESN's~\cite{JAEGER2007,LESNREVIEW2022,scardapane2017}, but ReLU was found to work better for our application. 

To investigate how ``trivial" it was to generate a classifier with the same capacity as our experiment, we generated 100 such LESN's at random, and found their average performance, and standard deviation. Hyper-parameters $a,\gamma,w_{\text{in}},\rho$, and sparsity ($p_s$) were tuned in sweeps to improve performance as much as possible for each width and depth, in order to give the digital reservoir a competitive chance. The reservoir's computational capacity varies by the number of shots.  Comparing Fig~\ref{Fig:Time-Independent}(b) to Fig~\ref{fig:si:lesn_spiral}, we found that, at around $10^3$ shots, our physical reservoir achieved a performance comparable to that of about that of a 32-dimensional LESN reservoir, as seen by the fact that both oscillate around 99$\%$ classification accuracy, within about one percent. In  Fig~\ref{fig:si:lesn_spiral}, a 64-dimensional reservoir was found to be enough to classify the spiral data points with perfect accuracy and a fairly wide choice of parameters. Our reservoir, then, achieved at \emph{least} the capacity of a 64-dimensional LESN reservoir, past around $5*10^3$ shots.

In the conventional view of reservoirs, the data must be sent into into a higher dimensional space where linear separability becomes possible~\cite{LESNREVIEW2022}. The dimensionality of our Hilbert space, given by the two dimensions of the qubit and the approximately 16 occupied levels of our storage resonator, limits the complex degrees of freedom we have available in encoding our data as the final, measured state. Consequently, we use this total Hilbert space dimension (times two due to complex amplitudes) as a proxy for the quantum resources used in terms of reservoir dimensionality. Given the roughly $2\times 16$ dimensions of Hilbert space used by our reservoir, achieving at \emph{least} the computational capacity of a 64-dimensional LESN reservoir is on the order of what would be expected for a large shot number. Indeed, the point of this comparison is to demonstrate that the computations performed by our reservoir cannot be trivially replicated by a digital reservoir with fewer resources with the same performance.

\putbib[%
 bibsm/general%
]

\end{bibunit}
\end{appendices}

\end{document}